# Improving Datacenter Utilization through Containerized Service-Based Architecture

Aos Mulahuwaish [a], Shane Korbel [b], and Basheer Qolomany [c]

[ab] *Department of Computer Science and Information Systems Saginaw Valley State University, 7400 Bay Rd, University Center, MI 48710 USA*

[c] *Department of Cyber Systems, College of Business & Technology, University of Nebraska at Kearney, Kearney, NE 68849 USA*

[a] *amulahuw@svsu.edu;* [b] *swkorbel@svsu.edu;* [c] *qolomanyb@unk.edu*

**Abstract**

The modern datacenter's computing capabilities have far outstripped the applications running within and have become a hidden cost of doing business due to how software is architected and deployed. Resources are over-allocated to monolithic applications that sit idle for large parts of the day. If applications were architected and deployed differently, shared services could be used for multiple applications as needed. When combined with powerful orchestration software, containerized microservices can both deploy and dynamically scale applications from very small to very large within moments—scaling the application not only across a single datacenter but across all datacenters where the application(s) are deployed.

   In this paper, we analyze data from an application(s) deployed both as a single monolithic codebase and as a containerized application using microservice-based architecture to calculate the performance and computing resource waste are both architected and deployed. A modern approach is offered as a solution as a path from how to go from a monolithic codebase to a more efficient, reliable, scalable, and less costly deployment model.

# 1. Introduction

Software advances of the 1980s through the early 2000s were primarily limited by the hardware that it ran on. The finite number of resources available on hosts meant that software was constrained or rate-limited in processing abilities. After the millennium turn, hardware advances started to surpass software's ability to consume all the available resources. This shift in dynamics has led to extra resources allocated to other software or where a single physical host could run many virtual hosts. Even with this deployment model, resources go unused and represent a significant opportunity for cost savings by modifying how we build and deploy software.

Datacenters and computer hardware have undergone several significant transformations over the last four decades. In the early '80s, mainframes took up entire rooms and performed specific tasks over and over very quickly. Mainframes at the time, and for the most part, remain specialized computing devices that are not widespread. Size and cost limited mainframes to large organizations to do specialized processing quickly. When hardware became smaller and less expensive during the '90s, a shift was seen to more generalized computing machines and the rise of distributed architecture [16], where datacenters were now comprised of hundreds or thousands of servers running parts of applications.

These smaller distributed computing systems grew more powerful at the turn of the millennium. The distributed servers became able to run more than a single application and eventually more than a single guest Operating System (OS), leading to a rise in virtualization of servers where many server instances may run on a single host. Virtualization led tech giants like Amazon, Microsoft, and Google to create large networks of datacenters across all geographies, starting the rise of infrastructure as a service as early as a decade ago. Infrastructure as a service has companies shifting away from using on-premises datacenters [17] to cloud-based solutions for hosting applications or hybrid on-premises/cloud deployments.

Software architecture over the decades has undergone an extensive series of transformations. In this paper, and to keep things focused, only changes that came about because of datacenter hardware advances will be explored. Early software was purposely built for processing bulk data over and over on mainframes. When distributed systems started to take over, and hardware was more generalized, the software could be designed to run at scale by adding more systems running the same software and then balancing each system's load.

In the last decade, two factors have altered the way software is architected: containerization and service-orientated architecture [1, 21, 22, 23]. Building software to run in containers is not a new concept. Containerization was created and has been around for decades to keep the kernel safe by isolating the running software kernel to take all the host's resources and compromise its stability. Keeping this critical interface operating makes containerization so appealing to prevent software that is not part of the operating system from affecting system stability. Containerization saw very little use until the Service-Oriented Architecture (SOA) [1] model was developed. SOA redefines a way to make software components reusable by defining service interfaces that can be used by other software. These interfaces utilize standardized communication so that they can be readily incorporated into new applications without the need to perform reintegration each time.

Each service in an SOA contains the code and data integrations required to execute an entire business function, such as looking up a destination on a map or distance between two given points. The service interfaces provide loose coupling, meaning they can be called without knowledge of how the integration is implemented. The services are exposed using standard network protocols—such as SOAP/HTTP [18] or JSON/HTTP [19]—to send requests to read or change data. The services are implemented in ways that enable developers to quickly reuse them to assemble new applications or functionality. In the SOA model, rather than having a single codebase that was deployed to a host, it is decomposed into a set of independent services deployed and designed to make service calls to one another to perform the application's total work. When applications are architecturally built to combine SOA and containerization in the modern datacenter, organizations are presented with the opportunity to improve applications' reliability and scalability while reducing the number of wasted resources by using containerized microservices. Microservices decompose a monolithic application through SOA before putting each microservice into a container.

Consider the following mock monolithic application, shown in Figure 1. This application comprises five different services (Authentication, Customer Frontend, Busines to Business Application Programming Interface (B2B Application Programming Interface), and a Database service).

In the monolithic application case, all services are contained within the same codebase, and to scale the application, a new instance of the application must be run on its host, as shown in Figure 1, by wrapping each service in a square box representing a physical machine, or a virtual machine. Breaking apart the application to move to an SOA model may start simply with taking the database service and removing that as part of the application and then having the application access the database via external calls from the application to the database.

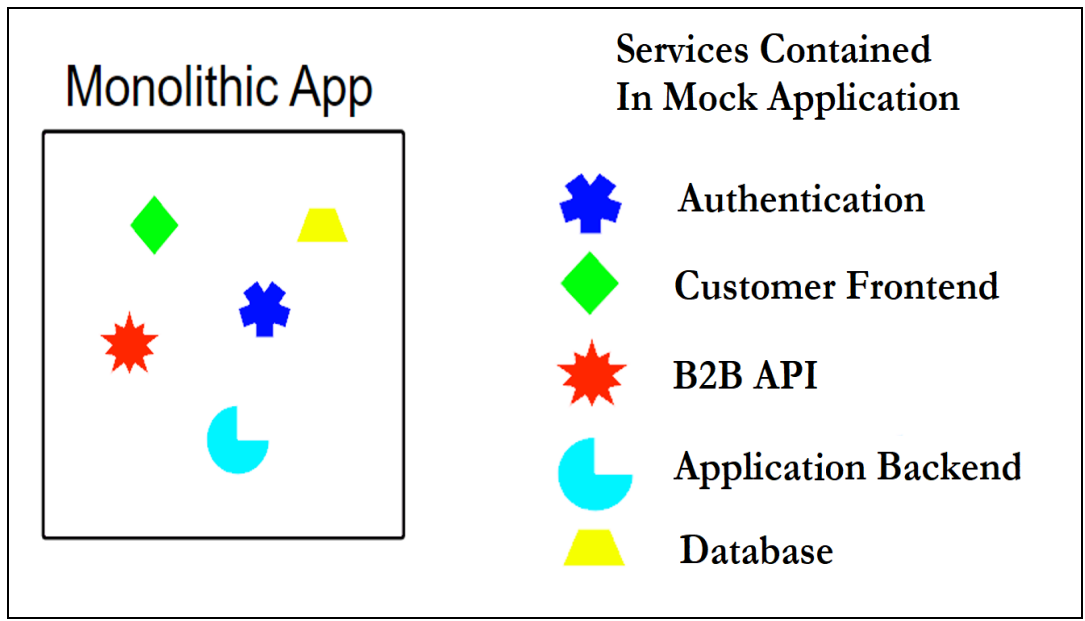

Figure 1: Mock monolithic application consisting of five services

The key difference here as an application starts to move from a monolithic to microservice-based architecture is that we can take the application and scale it differently from the database, which may have entirely different system requirements such as faster Input/Output (I/O) on the data storage layer. The application developers can then decompose all services from the application that make sense based on their functions, striving to develop highly cohesive and loosely coupled services for a proper microservice-based application. These new services can then be set up to scale horizontally as needed, increasing each service's number of instances as the demand for that service increases. If the host runs out of capacity to run more instances of each service, it becomes easy to add more nodes to the cluster running the microservices and create the new containers across multiple hosts. Figure 2 shows the transition in scaling between a monolithic application (left), a slightly refactored monolithic application (center), and an application composed only of microservices (right).

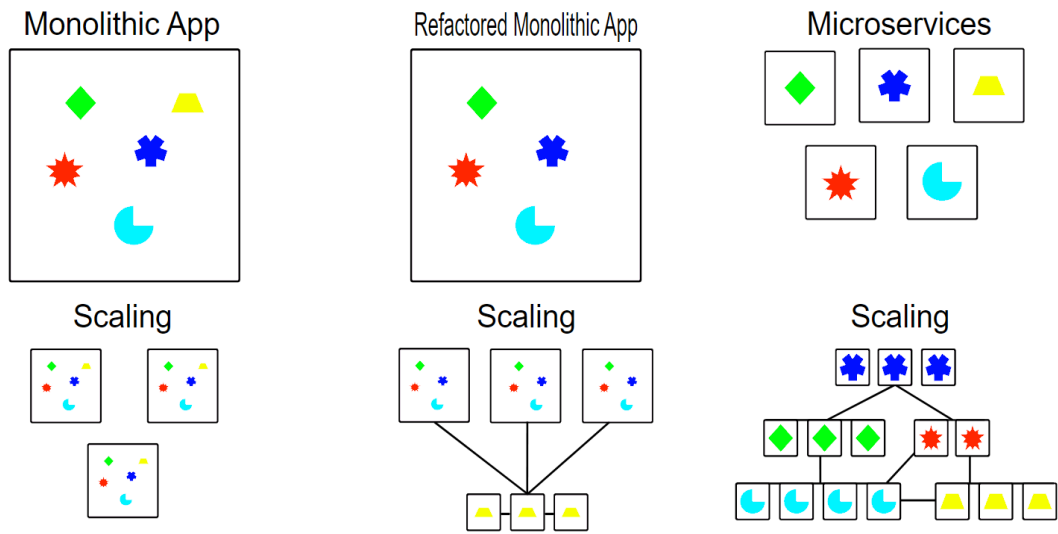

Figure 2: Difference in scaling between monolithic, refactored, and microservice software architectures (migrating from monolithic applications to microservices)

**1.2 Problem Definition**

Today's datacenters usually have many different hardware resources for running applications that typically get deployed from a single codebase. One type is large physical hosts running many virtual hosts. These large physical hosts have large amounts of Central Processing Units (CPUs), Random Access Memory (RAM), Graphics Processing Units (GPUs), storage devices, or any combination thereof. These hosts have modest to large amounts of onboard storage to accommodate applications that could have large amounts of disk I/O or tolerate the network latency of Storage Area Networks (SANs) that these hosts are attached to. While this storage is transparent to the hosts utilizing it as a physical disk on the machine and allowed for storage to be adjusted for those hosts dynamically, it comes at the cost of latency since the disks are not physically part of the host. These high-resource physical hosts then run many virtual hosts by allocating a set number of resources shared among all virtual hosts or dedicated to the virtual host. This flexibility of either having resources dedicated to a host or shared among all the virtual hosts allows for the over-allocation of the underlying physical host's total available resources to let an organization's operations team maintain all the hosts keep fewer computing resources sitting idle. Resource over-allocation presents the risk of running out of resources when demand across one or all the virtual hosts is high.

    The second type of host found in modern datacenters is a physical host running a single application without running any virtualization of a guest operating system on the host, making the host dedicated exclusively to the application being run upon it. The types of software applications that do well without virtualization may have a very high dedicated physical resource requirement, such as large enterprise databases, or do not handle network latency well. For now, these types of applications that require specialized hardware requirements and dedicated physical hosts will continue to exist. These systems will continue in their current states due to the necessity of having low latency for applications like databases performing hundreds to hundreds of thousands of requests quickly. Any latency introduced into

a system like that will have a substantial performance hit on the application. An example of this would be if a database used SANs for storage rather than using local solid-state storage. If the typical database call took 3ms using local storage, but when operating off SANs, the latency between the SANs and the database is 6ms, the architectural changes have increased the response time for that database transaction by double.

Underutilized Virtual Hosts

There is often much waste in applications running on virtual hosts with CPUs spending clock cycles keeping all the guest operating systems running even with over-allocating resources in virtual machines. RAM becomes utilized needlessly for non-application specific objects or used by the guest operating system being virtualized. Figure 3 shows graphs of several hosts' CPU utilization percentages. The space above each line represents unused resources. The CPU utilization is an average of everything running on each host, including the guest operating system and any application running. Each color represents a different node.

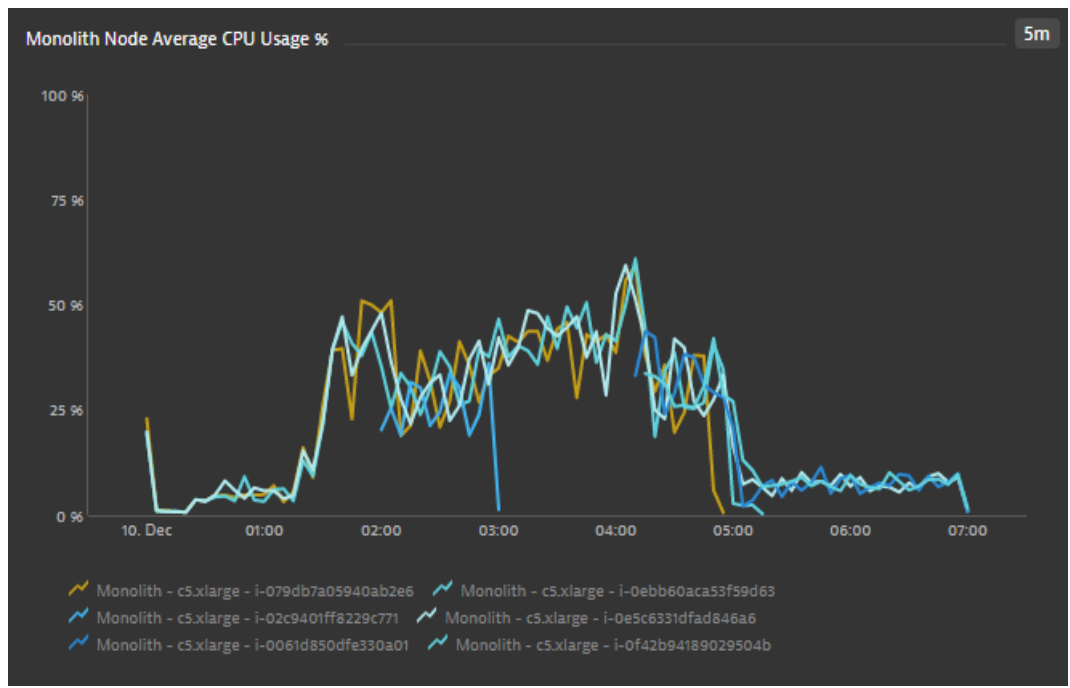

Figure 3: Graph showing average CPU utilization over time

These unused resources shown in Figure 3 represent a cost that cannot be recouped since any unused resources sit idle and are wasted. Depending on the number of underutilized hosts, this can represent high costs to the bottom line of an organization's operating costs that could be better managed using a containerized service-based architecture. The containerized system can then be managed by orchestration software and adding or removing nodes on demand. Re-architecture to containerized microservices would allow organizations to deploy many containerized applications to a small set of servers and run them just short of full load. Suppose the capacity total starts to affect user performance and another host is added for more resources. In that case, extra capacity can be quickly added, ensuring the application does not go down, and performance remains consistent. It also allows for rules to be set up to take that extra unit of capacity and terminate to save costs after the utilization returns to normal.

Monolithic Applications

To better utilize surplus hardware resources, how software is architected must be examined. Single codebase or "monolithic" applications currently represent most of those deployed in modern datacenters and are the first to change. Monolithic applications have several drawbacks that can be overcome or mitigated by microservice architecture. While simple to deploy, monolithic applications can snowball in complexity and be inefficient in their use of resources. Eventually, components within the code will become coupled in unforeseen ways, which can slow down changes and become more challenging to troubleshoot down the road should problems arise. The smallest change or update requires the entire application to be thoroughly tested and redeployed. Since the impact of changes is more challenging to measure across the entire codebase, testing the application becomes a laborious effort requiring an entire team of people. New technology and frameworks have a high cost of entry in terms of redesigning the entire codebase. Bugs in a single part of the application, such as a memory leak, can bring down the entire application and affect all hosts running the application.

**1.3 Contributions of this Work**

Many surveys, tutorials, and articles in existing similar works look at the same elements in this work and are broken down in detail in the next section. Where other works look at specific aspects, this work looks at the problem holistically while performing the testing at a scale approximating what would be encountered in a real-world production environment for a typical corporate application. Other approaches focused on looking at an application with only two different APIs. They called hundreds of transactions on small physical machines sized like consumer Network Attached Storage (NAS) hardware versus what would be found in real production environments for standardized physical or virtualized hardware. A NAS system is a storage device or system of devices connected to a network that allows storage and retrieval of data from a centralized location for users on a network. NAS systems are flexible and scalable, meaning that as the need for additional storage increases, you can add on to what you have. NAS is like having a local private cloud. It is faster, less expensive, and provides all the benefits of a public cloud locally.

    Most applications using an SOA model will have hundreds to thousands of APIs depending on their size and function. Network-attached storage hardware is consumer-grade SAN hardware. NAS hardware typically has a small amount of CPU and memory and a variable amount of space for hard drives to be added into a storage pool. This type of hardware is not robust enough for a testing representative of an application in the real world.

    What makes this work different from that found in the existing literature is the scope and scale. This work looks at many aspects of how monolithic single codebase applications and containerized service-based applications can differ, such as:

- Throughput of the applications (number of transactions) being processed over various user workflows, including scenarios such as login, searching, booking travel, calculation of recommendations, and checkout.
- Response time that these transactions take and the comparison between both architectures.
- Hardware utilization across all nodes for both architectures and how they operate and differ in scale.
- Finally, this work was based on being the most extensive scale of its kind across the software dimensions and in hardware with scaling. It is also the only work of its kind where the load testing and data collection was done completely autonomously through scripting and state of the art tools, with load testing resulting in peak transactions that simulate real user traffic generating over five thousand requests per minute to both the application frontends resulting in a peak number of service calls across frontend, backend and database services for both architectures being just under one million (peak value: 960 thousand) in a single five-minute interval as shown in Figure 4.

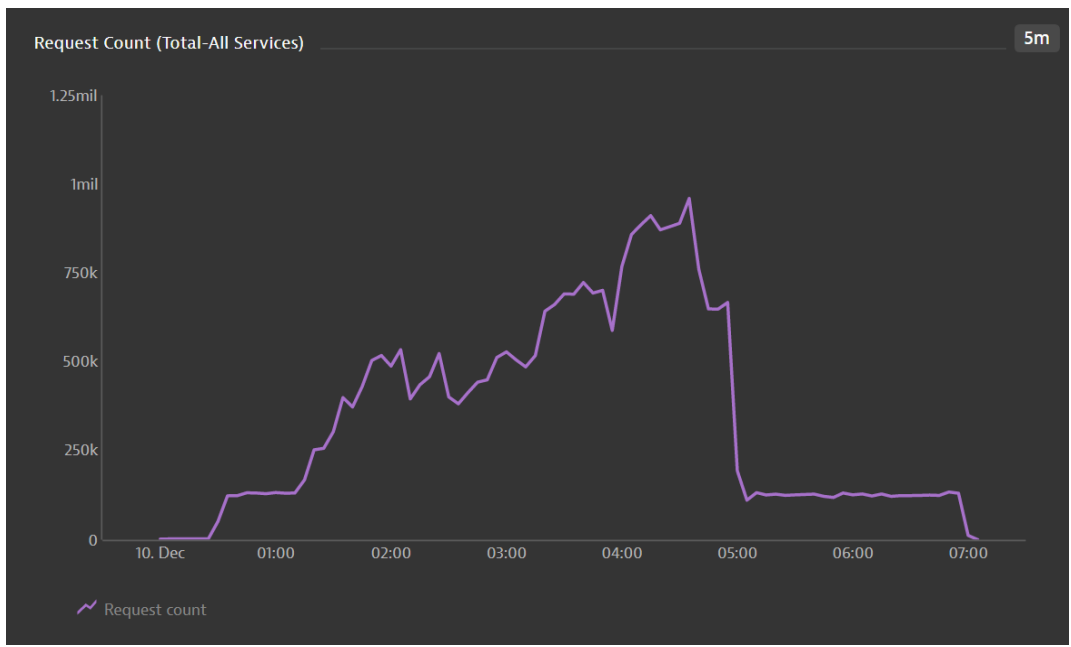

Figure 4: Total request count of all services, both monolithic and microservice architectures, during research testing timeframe

The remainder of this paper is structured as follows: Section 2 discusses the related work. Section 3 presents the research methodology and design. Section 4 presents the implementation and results. Finally, section 5 provides conclusions and future work.

## 2. Related Work

A handful of themes or approaches appear in computer science concerning literature about design patterns and breaking down a monolithic application into disparate services that may or may not involve containerization. Some studies that look at splitting monolithic applications apart look at coming up with a series of steps to follow. Additional studies are themed around if it makes sense to take a monolith apart at all. Other work [3] looks to understand the performance impact between an established monolithic application before and after it is split into separate services to look for any differences in the performance.

The work introduced by Gouigoux et al. [3] discusses the French company "MGDIS SA." This French company undertook to rewrite their core business software from a monolithic architecture to web-oriented architecture using microservices. They looked at the technical changes over three years and consumed 17,300 person-days to facilitate their applications' refactoring. The French company faced their applications being made obsolete and decided to start over from the drawing board. Many questions were raised by [3], such as how to refactor the monolithic application apart into different services and what parts should be rewritten alongside what technologies should be used. The services' granularity was also a critical thing explored since not every function or method of an application needs to be its own service unless it makes sense. Microservices running in containers should be loosely coupled with different or emerging tech stacks or technologies to improve the system.

In some cases, it was found beneficial to make specific functions not part of a microservice. This prevents becoming locked into a programming language or vendor. Those functions were taken serverless and implemented specific application parts to run on one-off serverless execution systems such as Microsoft Azure Functions or Amazon's Web Services (AWS) Lambda as the best path.

The findings in [3] pointed to one of the strengths of microservices-based approaches: to reuse elements that could be common. Using this design pattern, deduplicated elements are repeatedly designed for applications and then reused for multiple applications. An example of this would be a microservice for handling login or authentication of users among many applications. Lastly, the French organization found that specific microservices, once developed, could be turned into a revenue-producing stream for the organization by selling that bit of functionality to other organizations, which was not a business case that was considered when taking on the task of rewriting the core business applications. Performance increases were also observed after the applications were rewritten, which was impossible in the old architecture despite efforts to address those issues. The work concluded that the organization would see a return on investment in less than five years, even though that time had not passed at the time of the paper. The estimated lifecycle of the applications rewritten was set at being ten years. Given it was being written in a microservice-based architecture, it could also be hypothesized that iterations and maintenance could extend that application's life expectancy compared to monolithic approaches. Adding new functionality at a fractional cost would be more trivial than doing so for an extensive monolithic rewrite like the one the French company had just undertaken.

Villamizar et al. [7] compare the response time between an application deployed as a monolithic design approach and a microservice-based system where both were deployed to comparable Amazon Web Services Elastic Compute Cloud (AWS EC2) instances. Dedicated resources are more expensive since they cannot be used for a pool of virtual machines like shared resources; they are instead dedicated to the specific instance. The application deployed in work had only two services (S1 and S2). The case study introduced by the authors recognized that while most enterprise applications would be composed of many more than just two services, only two services were custom created with specific performance built into the service to reduce the study's complexity. The service S1 implemented a CPU-intensive algorithm to generate every payment plan, and the typical response time was around 3000 milliseconds. For the second service, the typical response time was around 300 milliseconds and used the database.

The authors used JMeter [8] to configure a workload to load against each service for both design pattern deployments. Using this constant load against the application, they measured the performance and then figured out the right size EC2 instance in AWS for both the monolith and microservice-based application versions. This correct sizing was for them to handle the load required, which showed a cost savings of the microservice-based deployment of $70.56, representing a savings of 17%. The application's average response time for S1 in the monolithic deployment was 2837ms, and for the microservice-based deployment was 3229ms. This showed a slight performance difference, but the cost benefits outweighed this slight performance hit.

Flygare et al. [9] discuss how the services perform under different loads and look at the host's different architectures' resource consumption differences. Independent variables such as the number of users simulated and dependent variables such as RAM and CPU usage break apart what is running on the system from the system itself since the system resources are static, whereas the load is variable. It keeps the load generated against those into sets and then creates a data matrix that compares all those aspects to break down the performance aspects. While this approach was adequate in comparing the various aspects involved, it was limited in the number of configurations from the hardware perspective. The authors failed to display the more effortless scalability of a containerized microservice versus monolithic architecture. For instance, they used low-end physical Intel NUC (Next Unit of Computing – small form-factor sized computers) computers for their hardware versus something more standardized like an Amazon EC2 instance for the tests. Their conclusions were like those in [7] that showed monolithic architecture was faster regarding how the application performed when large numbers of concurrent transactions are happening; however, they did not scale the application hosts as the load increased in both papers.

Saransig et al. [10] looked at how the total completion time affected both monolithic and containerized microservices' performance cases. The work reports the total time to complete all the transactions and then breaks down the number of requests per second each type of architecture could handle. It was concluded that regarding host resources consumed, monolithic architecture outperformed containerized microservices. However, while efficient in resources consumed, they did not outperform in all cases regarding application response time. So, while the hardware used remained the same, the monolithic architecture was slower overall. It concluded that the performance gains outweighed marginal extra resources consumed. Performance is far more an ideal scenario in the real world. High concurrent users require higher application throughput to deliver the shortest response time possible when resource limitations on the host system are rarely considered in the modern datacenter where additional resources can be added quickly on demand.

The author in [11] takes things further than previously mentioned techniques by building a typical social media application in monolithic and containerized service-oriented architectures. As load increases, it scales each service at specific points and reports back metrics on how long it takes to add additional capacity. The work has a decent representation of services that do tasks such as login, registration, retrieving a newsfeed, and user-to-user messages. It then did what many other works have done by running predetermined amounts of transactions of all different types against both architectures and reporting back on response time and failure rate. There was no breakdown of what sort of hardware was used for either deployment architecture. The work talks about scaling the applications for both different architectures; it does not say if this is done manually or automatically triggered. No information is given on how many starting or total instances of each were represented in the data. Without including that information, it is impossible to form a complete picture of the application's performance since no hardware resources are consumed for either architecture.

## 3. Research Methodology and Design

This section explores the decisions around the different technologies available and why one is chosen over another in the overall design. The main aim is to present the reader with a methodology and design that improves the scale and scope beyond any previous works done, such as increasing the number of different transactions per microservice, the number of microservices, and finally, the scaling of the applications and hardware they run on.

### 3.1 Proposed Methodology

Several avenues can be taken to implement a containerized service-based application that entirely takes advantage of all the benefits this architecture has to offer [24, 25, 26]. From a pure research perspective, the most exciting way to implement it while doing new research that does not currently exist would be to take an existing monolithic application that has several distinctly different services or functions and then working on splitting those into microservices or smaller application components that utilize APIs and rest interfaces to communicate to each other like what was done in [12]. Unlike [12], however, the design improves the scale and scope beyond any previous works done, such as increasing the number of different transactions per microservice, the number of microservices, and finally, the scaling of the applications and hardware run on. Once the monolithic single codebase application has been converted, the analysis would be done to test the performance and deployment benefits. Similar outcomes can also be done with an application that has already been converted to a microservice-based architecture from a monolithic architecture. This approach takes less time since the application will already exist in both a monolithic and microservice architecture. This paper will utilize the latter. An example application built to be deployed as a monolithic architecture and a containerized microservice-based application is "EasyTravel," a demo travel booking application. "EasyTravel" is explained in detail in Section 4 when detailing the implementation.

To analyze the performance of EasyTravel as both a monolith and a containerized microservice, EasyTravel will be deployed both ways in AWS on EC2 instances that are identical from hardware and scaling perspectives. Both will be configured to automatically scale the number of nodes when a CPU saturation on the host reaches 50% on average. When an individual service's container reaches 80% CPU utilization in the microservice-based application, another container of the same type will be deployed. Another set of servers will be set up to produce synthetic use on both deployed applications to simulate load. These two sets of servers will be scaled programmatically. At set intervals, additional stress will be applied to both the monolithic and microservice-based architectures. They will also test against different application aspects to examine the container orchestration tool's scaling benefits in different scenarios. These scenarios include loading the homepage, logging in, searching for travel arrangements, and booking these travel arrangements through a simulated checkout. This testing will analyze the performance benefits of containerization over monolithic applications' dynamic, scalable environment. It also measures the load on the host and the application's performance by looking at the response times of the various functions performed on the application.

## 3.2 Design

The research work was designed and performed to explain the microservice architecture and containerization/orchestration as a powerful alternative to monolithic architecture focused around Docker containers and Kubernetes orchestration on AWS elastic Kubernetes service. This paper will focus on how architecture affects capacity, scalability, response time, and key performance indicators. It will also identify each type of application architecture's strengths and weaknesses and the different complexity levels in creating, deploying, and maintaining these applications. This will be done by:

1. It was finding or developing a mock monolithic application that could be converted to a containerized microservice architecture. "EasyTravel" was selected for this purpose.
2. Deploy AWS in monolithic and microservice architectures under identical underlying application hosting infrastructure.
3. Collect utilization and performance resources and then analyze the ability to handle load tests in monolithic and microservice architecture forms to measure how architecture affects capacity, scalability, response time, and other key performance indicators.

# 4. Implementation and Results

This section explains how the implementation was set up inside an environment that mirrored what a real production level environment in AWS would look like. After the implementation has been explained, how the testing on the implementation findings will be presented. Some of these findings are exactly as predicted, while some were unforeseen, and all of these will be explored in detail in Section 4.2.

## 4.1 Implementation

The implementation of this paper can be broken down into three distinct parts: the first part is EasyTravel, the application used, which can be deployed in monolithic or microservice containerized architectures. The second part is the datacenter implementation done in AWS to host both the EasyTravel application and its load generating counterpart while providing all the necessary tools to carry out completely automated deployment scaling of the application. Finally, bash scripts and tools such as Dynatrace [13] are used to facilitate the automated tests and monitoring of the application to provide back the data and performance of both deployments of the application.

EasyTravel (Figure 5) is a fictional travel booking website that provides a web portal that allows users to log in, search for journeys to various destinations, select promotional journeys directly offered, and book a journey using credit card details. Additionally, a Business-to-Business (B2B) web portal for travel agencies is provided where travel agencies can manage the journeys they offer and review reports about made bookings.

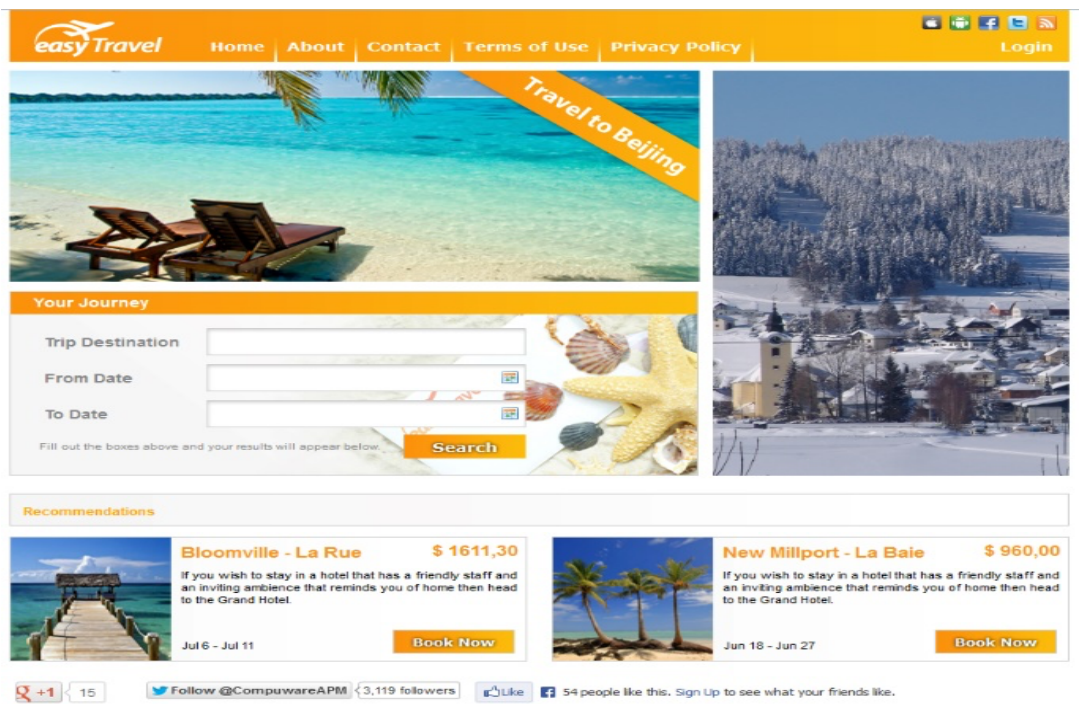

Figure 5: EasyTravel User Interface (UI)

All services for the monolithic style architecture are installed from a single installer. In contrast, the application's microservice-based version is broken down into the following services that run in their respective containers: frontend, backend, loadgen, and database. The application's front end is where users will interface with the application to simulate booking travel within the application. The backend process facilitates all the required processing of those bookings and interfaces with the database tier that holds the user booking data. While there are only three containers for EasyTravel when deployed as a microservice, several sub-services are in each of those containers, as shown in Figure 6. In addition to the frontend, backend, and database tiers, there is also a load generation component to the application to simulate how users would use EasyTravel. For this work, the load generation is turned off on the monolith and microservice clusters' nodes. This is done because the load generation will be done from different clusters, not to load the hosts on the application. However, this feature is used on the load generating node clusters. The application is modified to access either the monolithic or microservice cluster's application load balancer on those nodes rather than the default, localhost. For a visual reference, please see Figure 7.

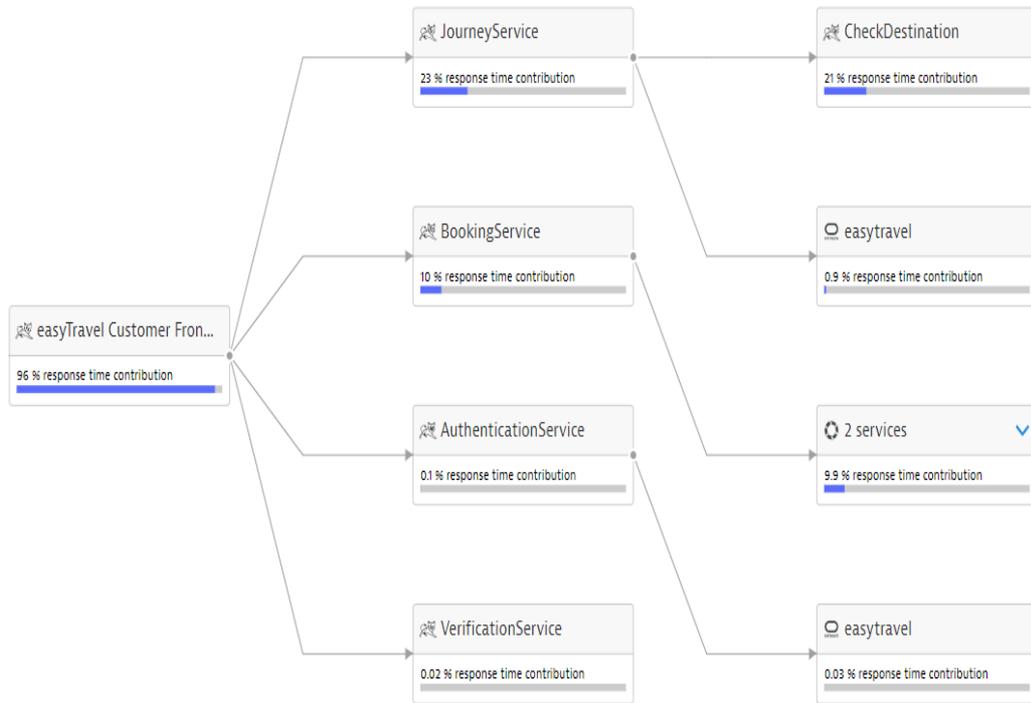

Figure 6: EasyTravel interrelated services

All application deployment was done on AWS and monitored by Dynatrace. Dynatrace [13] is a software-intelligence monitoring platform that simplifies enterprise cloud complexity and accelerates digital transformation. With Davis (the Dynatrace Artificial Intelligence (AI) causation engine) and complete automation, the Dynatrace all-in-one platform provides answers, not just data, about the performance of applications, their underlying infrastructure, and end-users experience. Dynatrace modernizes and automates enterprise cloud operations, releases higher-quality software faster, and delivers optimum digital experiences. Dynatrace seamlessly brings infrastructure and cloud, application performance, and digital experience monitoring into an all-in-one, automated solution powered by artificial intelligence. Dynatrace provided several key capabilities in monitoring and providing metrics back about the tests run for this work, including real user monitoring, server-side service monitoring, network, process, host monitoring, cloud and virtual machine monitoring, and container monitoring. Real user monitoring analyzes the performance of all user interactions with the application. Real user monitoring also enables application availability monitoring, verification of correct display of User Interface (UI) elements, third-party content provider performance analysis, backend service performance analysis (down to the code level), and performance analysis of all underlying infrastructure. Web applications consist of web pages served by web servers (for example, Apache Tomcat) and web containers (for example, Docker). The web requests that are sent to a specific Tomcat server are an example of a server-side service. Server-side services may be of various types like web services, web containers, database requests, and custom services. Dynatrace OneAgent can provide details about which applications or services interact with other services and which services or databases a specific service call. Dynatrace enables monitoring of the entire infrastructure, including hosts, processes, and networks. Dynatrace enables log monitoring and view information such as the network's total traffic, the CPU usage of hosts, the response time of processes, and more.

Dynatrace OneAgent monitors the entire stack, including private, public, and hybrid cloud environments. Dynatrace OneAgent auto-detects all virtualized components and keeps up with all changes. Dynatrace OneAgent can be integrated with virtualized infrastructure, the processes that run on them, and the applications. Dynatrace seamlessly integrates with existing Docker environments and automatically monitors containerized applications and services. Dynatrace hooks into containers and provides code for injecting OneAgent into containerized processes. There is no need to modify Docker images, modify run commands, or create additional containers to enable Docker monitoring. Dynatrace automatically detects the creation and termination of containers and monitors the applications and services contained within those containers.

An EC2 instance was created using 64-bit architecture and running Ubuntu 18.04 LTS as its operating system. It was then configured as a template to be used later to deploy the EasyTravel nodes as a monolithic style application. This instance was also modified slightly to be used in the load generation nodes. This EC2 instance had all the required scripts that were created to download EasyTravel and Dynatrace, install all pre-required components, install Dynatrace, install EasyTravel, copy configuration files used to alter the behavior of EasyTravel based on the node it was running on, and finally to start the EasyTravel application via CRON [14] upon the startup of the EC2 instance. The software utility CRON, also known as a CRON job, is a time-based job scheduler in Unix-like computer operating systems. Users that set up and maintain software environments use CRON to schedule jobs to run periodically at fixed times, dates, or intervals.

Finally, this EC2 instance was saved as an Amazon Machine Image (AMI) for deploying all the monolithic cluster nodes and then modified slightly to be used as the monolithic and microservice load generation nodes. An AMI is just a template that can be used to create virtual machines. Each time the node came online, it would start generating traffic to the load balancer of either the monolith or the microservice cluster based on its configuration. The load generation started at a set amount and ramped up slightly over the first few minutes after the node came online. For the load generation, several flows are used, including loading the homepage, calculating travel recommendations, showing special offers, finding journeys, finding locations, authenticating a user, and booking travel. The traffic is set to simulate real users and, therefore, has a bit of variation to it; however, throughout tens to hundreds of thousands of requests per minute during testing, this variation does not impact the results in any meaningful way. For the microservice containerized version of EasyTravel, the application was deployed to an Elastic Kubernetes Service (EKS) cluster running Kubernetes version 1.17. The deployment file for EasyTravel to run on EKS cluster. Each of the different microservices had different millicore CPU requests assigned in line with each container's expected load. In Kubernetes, limits and requests for CPU resources are measured in CPU units. [15] In Kubernetes, one CPU is equivalent to 1 vCPU/Core for cloud providers and one hyper thread on bare-metal Intel processors. Fractional requests are allowed. A container with spec.containers[].resources.requests.cpu of 0.5 is guaranteed half as much CPU as one that asks for 1 CPU.

The expression 0.1 is equivalent to the expression 100m, which can be read as "one hundred millicpu." Some people say, "one hundred millicores," which is understood to mean the same thing. A request with a decimal point, like 0.1, is converted to 100m by the application programming interface, and precision finer than 1m is not allowed. For this reason, the form 100m might be preferred. CPU is always requested as an absolute quantity, never as a relative quantity; 0.1 is the same amount of CPU on a single-core, dual-core, or 48-core machine. The way pods (single or multiple instances of containers of the same type) with resource requests are scheduled is as follows: when the Pod is created, the Kubernetes scheduler selects a node for the Pod to run on. Each node has a maximum capacity for each resource type: the amount of CPU and memory it can provide for Pods. The scheduler ensures that, for each resource type, the sum of the resource requests of the scheduled containers is less than the node's capacity. Although actual memory or CPU resource usage on nodes is very low, the scheduler still refuses to place a Pod on a node if the capacity check fails. This protects against a resource shortage on a node when resource usage later increases, for example, during a daily peak in request rate.

Since the frontend ended up being the most utilized microservice in initial testing, the reservation request of 200 millicores was requested for each frontend Pod. Remember that Pod will be created on any node in the cluster with at least that amount of CPU available otherwise, the Pod will be scheduled to run but not be deployed if the capacity is not available. In the testing conducted for this work, it was never a situation that was run into since another node of capacity was configured to be created should the cluster nodes' overall utilization exceed 50% CPU utilization. That is to say before the cluster as a whole would run out of capacity for starting another instance of the container, another node would be added to the cluster, thereby expanding the capacity beyond the requirements to start the container. The backend microservice and database were both set with a reservation of 400 millicores. The Pods' initial configuration was to have two frontend microservice Pods, two backend microservice Pods, and one database microservice Pod. Once deployed to EKS, the Pods' horizontal autoscaling was configured to try and keep the per Pod CPU utilization below 80%. Once the threshold was violated, another Pod of that same type was created. This autoscaling of Pods within the nodes' autoscaling creates a robust system able to handle changing workloads. Autoscaling can respond to surges in requests by offering another vector to scale and handle more threads per server for requests without changing application parameters or the underlying code. Using knowledge of the maximum a Pod can handle, more Pods can be deterministically created by software monitoring one of several metrics such as CPU/memory utilization, network, or disk I/O, or even the number of requests per second reduces the number of Pods as demand reduces. Scaling frees up resources on worker nodes to be used for other applications creating a smaller overall footprint, utilizing overall resources better, and adding or removing the capacity for Pods on demand by just adding or removing additional cluster nodes.

Both the monolithic and microservice clusters were set to auto scale and add an EC2 instance node to the respective cluster once the nodes' overall CPU utilization exceeded 50%. This metric was chosen since CPU saturation was the only constraint ever exceeded in testing different load generation scenarios. It is set to be quite low at 50% since adding in nodes for the monolith required considerable lead time to startup. This low value was due to each new node having to be provisioned, started within AWS, downloaded all dependencies and software, and only then could it handle requests with the other cluster nodes. This high lead time to start a new node for the monolithic architecture meant that if the CPU threshold for adding a new node was too high, CPU exhaustion on the other nodes could lead to failures. When this happened, it would lead to retries from the load balancers, which made even more requests, which would then fail, and once this happened, the monolithic architecture cluster may never recover no matter how many nodes were added. The microservice architecture never had this issue in initial testing due to the ability to react through autoscaling quickly; however, for the sake of keeping the underlying cluster parameters identical, the same was configured for both. Neither memory, disk I/O, or network saturation ended up as the bottleneck for the application in either architecture or deployment methodology.

Load generation was done identically for both the monolithic and microservice-based architectures, with only the load balancer target Uniform Resource Locator (URL) being changed. Load generation for each cluster was done from AWS T3 medium instances that were set to launch using the load generation image that was created, which pointed the node to the AWS application load balancer of the EKS cluster running the microservice container-based version of EasyTravel. AWS T3 medium instances were also used for the monolithic architecture load generation and configured to use those nodes' AWS application load balancer. Cluster nodes for both the monolithic architecture and the microservice container-based architecture ran on AWS EC2 C5 Extra Large (XL). The resource sizing values of the instances used at the time of this writing can be found in Table 1.

Table 1: Overview of AWS instance resource sizing used in testing

| Instance   | vCPU | Mem (GiB) | Storage  | Network (Gbps) | On-Demand Linux Usage Cost |
|------------|------|-----------|----------|----------------|----------------------------|
| t3.medium  | 2    | 4         | EBS-Only | Up to 5        | $0.0416 per Hour           |
| c5.xlarge  | 4    | 8         | EBS-Only | Up to 10       | $0.17 per Hour             |

After the implementation was complete, the testing was done in three stages and was completely autonomous. The first stage was to set up all environments for testing. Control over the load generation was done from an EC2 instance outside of any of the four clusters EC2 instances and will be referred to as the admin node. The admin node had permissions to call AWS APIs to adjust the number of running instances for each autoscaling cluster. A diagram of the setup implementation with all the different clusters, load balancers, instance types, number of instances, and the admin node can be found in Figure 7.

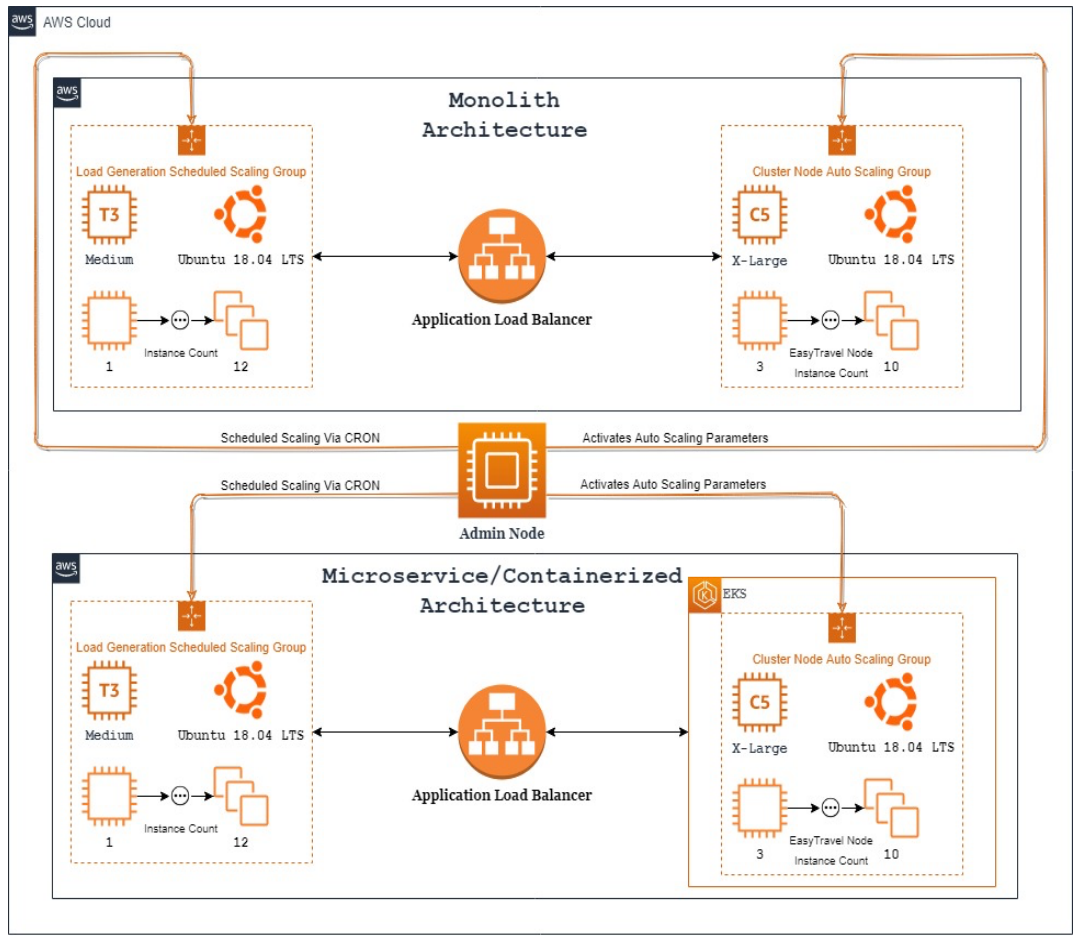

Figure 7: Architecture overview of the implementation

Scripts were created to facilitate all changes made to the environments as needed for the testing, such as increasing node counts for the load generation clusters and setting the min/max/desired number of application architecture cluster nodes. The scripts were then added to a CRON job on the admin EC2 node and ran on a pre-determined schedule. At 00:00ct on December the 10th, 2020, the initialization script was initialized that set the desired and minimum number of nodes in both the monolithic and microservice clusters to three EC2 instances of size C5 XL while also setting the maximum nodes to ten. This warm start was done to ensure that the nodes came up as expected and had time for some of the synthetic transactions from the data collection and monitoring tool Dynatrace to confirm everything looked as expected. At 00:30ct on December the 10th, 2020, stage two started by adding in load generation initiated from the admin node, and the CRON job set the min/max/desired number of load generation node count to one for each application architecture type to start putting the load on both deployment architectures of the application. This state was maintained for forty-five minutes to build a baseline for each application that could be compared and show any variation over time of the load generated against each application. At 01:15ct and every fifteen minutes after that, the admin node used CRON to call scripts to increase the number of min/max/desired load generation nodes by one until there were fifteen total load generator nodes generating traffic against both application architectures via their application load balancers with the final load generation nodes being added at 04:30ct. The number of nodes, fifteen, was to ensure that a significant amount of load was generated over quite a long duration at realistic intervals. At 05:00ct, stage three began where the admin node set the min/max/desired number of load generation nodes for each application architecture back to one node again instead of stopping every load generation node at once. This was done to demonstrate how the two architecture clusters would bleed off any extra nodes created because the cluster average CPU per node utilization percentage violation of 50% that may have occurred returns everything back to the baseline. For both architectures, this single load generation node continued until 7:00ct when the admin node terminated all load generation nodes, the microservice cluster nodes, and the monolith cluster nodes.

### 4.2 Results

The proposed methodology was designed to take an application functionally the same but architected in both a monolithic and microservice containerized way to compare the application's function and performance across a wide range of performance dimensions. This was then analyzed to show how the underlying host utilization is better utilized when components of an application are broken apart into services and containerized to be scaled up on the same host to add in additional capacity under higher workloads. In monolithic architecture, if the application function is to book travel arrangements, it cannot just create additional capacity if the login service is in high demand without deploying the entirety of the application to another host.

During the implementation, the expected result of which architecture performed better turned out to be the containerized microservice version of the application; however, it was not a bottleneck that was expected nor foreseen when planning the testing implementation. It was hypothesized that due to the more serialized nature of the monolithic architecture where user workflows occur, eventually, under the right load, one of those services would no longer keep up with demand. Which of the components eventually become the bottleneck was unknown. The workflow on each host at a high level is a frontend that the user interfaces with a backend that processes travel arrangements and talks to the database where all the user, destination, booking, and other stateful information are stored. Whichever bottleneck occurred, it was expected that the underlying host would be CPU bound first since the smaller scale testing suggested the application, once it reached a certain point, became CPU bound, slowing down transactions and making the application unresponsive. For that reason, the AWS application load balancers, the autoscaling groups for both architectures, would be based on average CPU utilization, as will be discussed in detail later. Due to how the monolithic architecture ended up handling garbage collection in the extensive scale testing, the application would end up getting into a state where it would become severely degraded and not exceed the average threshold value that was set up for both application architectures to try and keep the overall cluster utilization at fifty percent.

Consideration was given to using different autoscaling metrics such as network bytes in or request counts per target. However, these metrics would have scaled the microservice cluster in ways that would not prove out the intent or the purpose that this paper was aiming to provide. Consideration was also given and attempted at changing the instance sizes of the AWS EC2 that both architectures were run upon. A test was performed with instances that were using half as many CPUs as outlined in Section 3.1 Proposed Methodology, and this did not change the outcome since the issues with garbage collection would have occurred the same way given any number of cores per host.

In the end, while the scaling elements of the methodology did not go as expected for the monolithic architecture, the results and insights gained from the testing are no less significant to the core thesis that monolithic architecture presents challenges when trying to utilize resources under an unpredictable load test.

This section evaluates how containerized microservice architecture better utilizes resources over monolithic architectures; there are many dimensions to consider. This section will explore some of these many different metrics gathered around the performance of both the architectures and the host each application is running on and highlight each type of architecture's key advantages or disadvantages regarding these metrics.

A. Throughput

Throughput represents the amount of work, typically represented as the number of requests that can be fulfilled per instance per host. As the number of requests grows closer to the maximum throughput of the instance, response time will likely increase alongside queuing, or errors can occur, rendering the application unusable. Throughput is also crucial because if requests start getting rejected or queuing takes place, then timeouts can occur. Timeouts result in retries, which exacerbates the problem by flooding in even more requests as users or other systems keep trying to perform a failed request until it works, effectively resulting in an unintentional distributed denial of service attack.

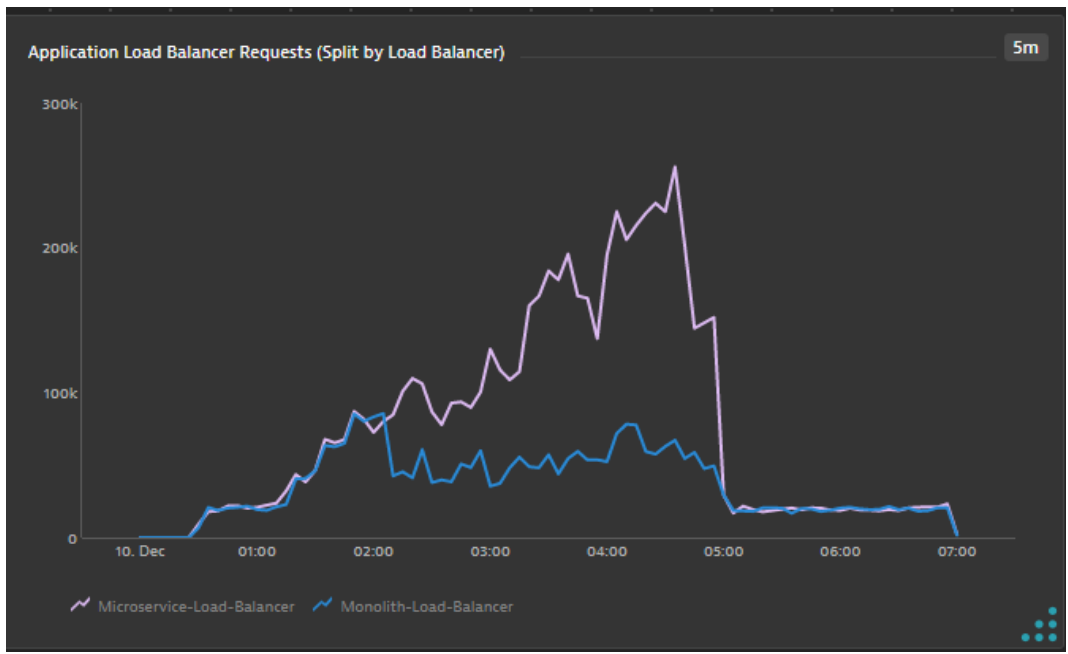

Figure 8: Application load balancer requests (split by load balancer)

Figure 8 shows that up until 2 am during the test, the number of requests being served by both application architectures remained nearly identical. At 2 am, when the 5th load generation node was added, a divergence occurred where the monolithic architecture was no longer able to keep up with the total number of requests coming in until 5 am when the load generation nodes were reduced to one per architecture, and then the throughput of both architectures reconverged once again. This also validates that there is no meaningful difference in the number of requests being generated from the load generation clusters to their respective application architectures as mentioned in Section 3.1 since the load generation was designed to simulate real users and had some variance in what type of requests were performed each time the load generator made a new request.

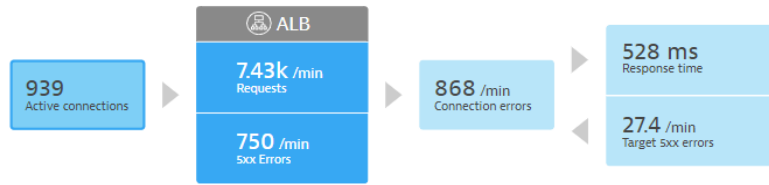
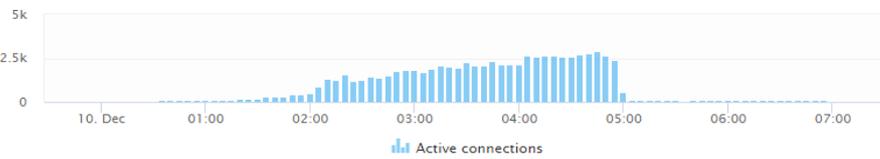
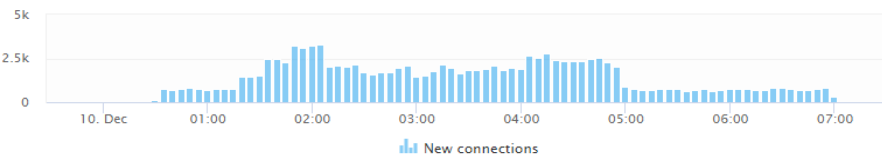

Figure 9: Detailed breakdown of monolithic load balancer activity

The detailed breakdown of load balancer activity in Figure 9 shows that the monolithic architecture has more active connections but fewer new connections than its microservice counterpart, as shown in Figure 10. This demonstrates how, when the monolithic architecture reaches its maximum throughput, it cannot take any new connections since it tries to fulfill the current requests from the already active connections.

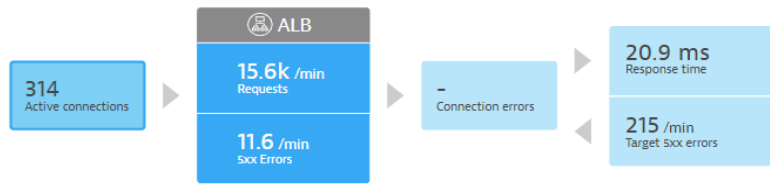

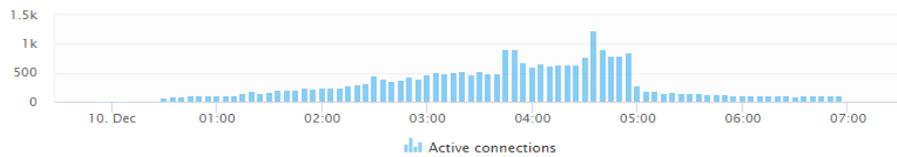

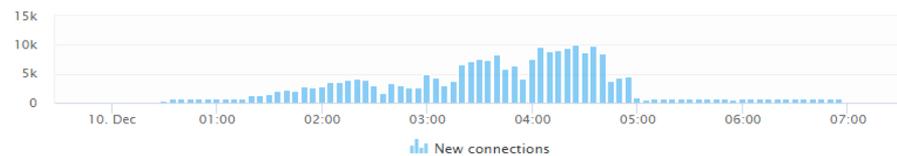

Figure 10: Detailed breakdown of microservice load balancer activity

These test results show the opposite for the microservice architecture in Figure 10. This results from fulfilling requests by adding additional capacity in the form of additional container instances as needed through rules designed to bring more container instances online on any of the services nearing their maximum throughput. The number of requests per minute for the microservice architecture is nearly double that of the monolithic architecture.

B. <u>Response Time</u>

Response time represents the amount of time it takes to fulfill a request from when the request was initiated until it was completed and returned successfully. Many different factors can affect the outcome of application response time, such as the device that initiates the request, the network latency between the client/server, the capacity of the infrastructure serving the request, and many more.

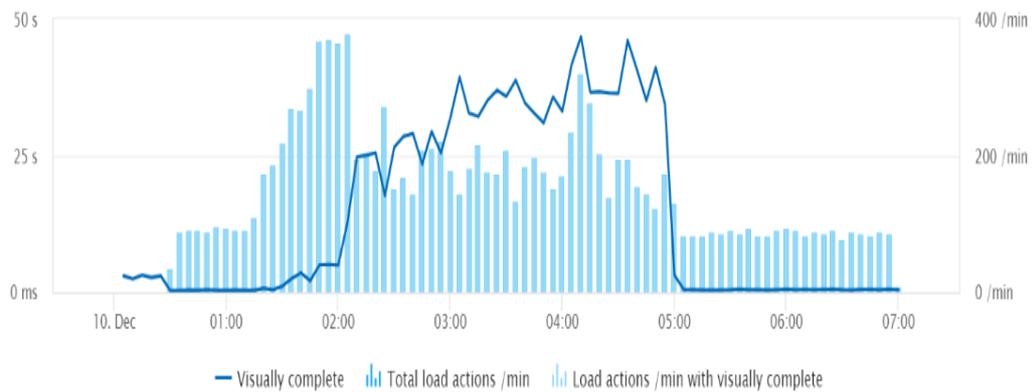

Figure 11: Monolith response time vs. the number of requests

Looking first at the Monolithic architecture overall in Figure 11, the response time spikes up substantially once, reaching about 400 requests per minute when the fifth load generation instance was activated. Response time got so high that the number of successful requests dropped considerably. The monolithic architecture could not keep pace with the load being generated against it—the peak response time when load generation at its highest resulted in requests taking nearly 50 seconds.

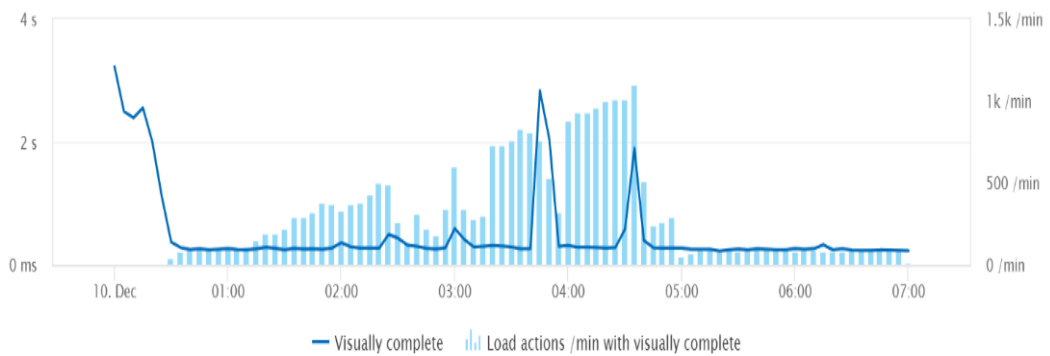

Figure 12: Microservice response time vs. the number of requests

The microservice architecture handled load increases much better in comparison. Figure 12 shows a few dips in the total throughput of the microservice architecture above. However, it responded to the increasing load before reaching a point where response time dramatically increased. When response time did increase, the microservice-based architecture quickly added units of capacity in the form of more replicas of the service under the most load. Despite serving nearly 4x the number of requests under peak load, the microservice architecture's response time average never exceeded four seconds. The average overall was around 300ms compared to the monolithic, which spiked up to around 50 seconds.

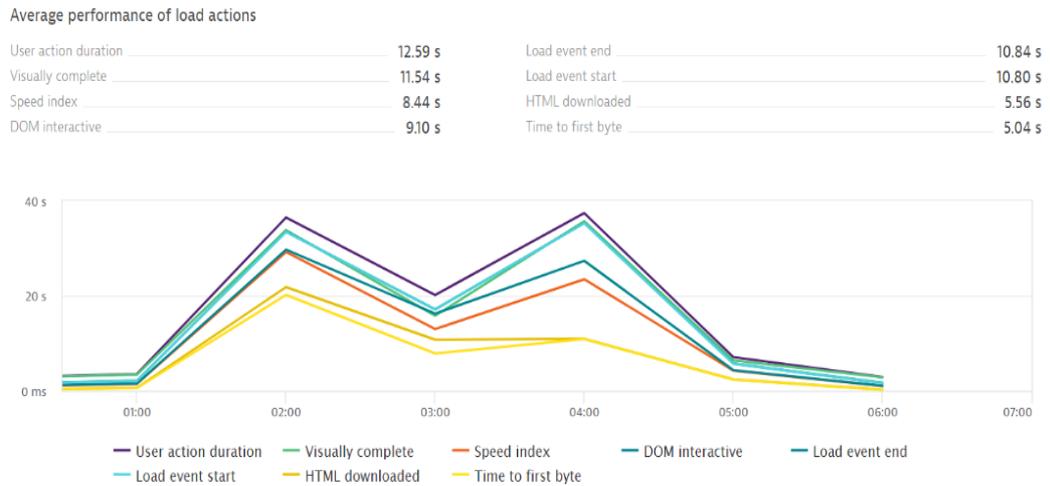

Figure 13: Synthetic monitor performance, monolithic architecture

In addition to the load generation traffic being produced on the application architecture, a synthetic monitor was set up to test the application from locations outside of the AWS cloud to measure the response time as if a real user were accessing the application from different parts of the USA. These tests were set up to run from Oregon, Chicago, Cheyenne, Los Angeles, South Carolina, and Texas. As shown in Figure 13, this external monitoring validates the results gathered from monitoring agents installed onto the application hosts and shows the same spike in response time around 2 am and returning to average around 5 am for the monolithic architecture. These synthetic monitors are the equivalent of a simulated user visiting the application using a web browser. They record response time and availability along with wc3 timings and the load times of each element on a given URL. These synthetic monitors can also simulate user input and be multi-step, but only single URL monitors were used for this work.

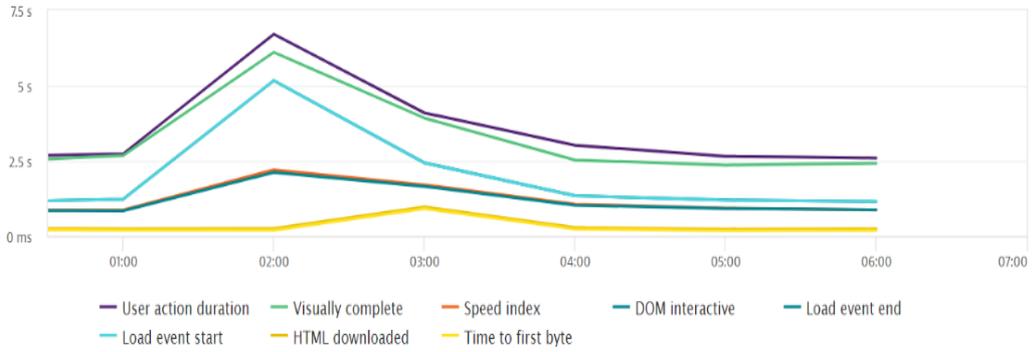

Figure 14: Synthetic monitor performance, microservice architecture

The microservice architecture agent-based monitoring results were also validated using synthetic external requests. A small spike in response time was observed around 2 am as the requests increased but quickly returned to a more normal baseline as more microservice replicas were added, shown in Figure 14.

C. Errors

While many types of errors can occur in an application while being used, the testing focused on bad requests (HTTP 4xx errors) and server errors (HTTP 5xx errors). 4xx errors indicate that the server did respond to a request. However, it will not process the request, usually due to something client-side, such as the request being malformed and invalid syntax, among others. 5xx errors indicate that the server is unavailable to process the request at all. 5xx errors are more severe than a 4xx error since 4xx errors may be retried and successful. In contrast, a 5xx error typically means the system is partially or entirely down and unable to process the transaction at all.

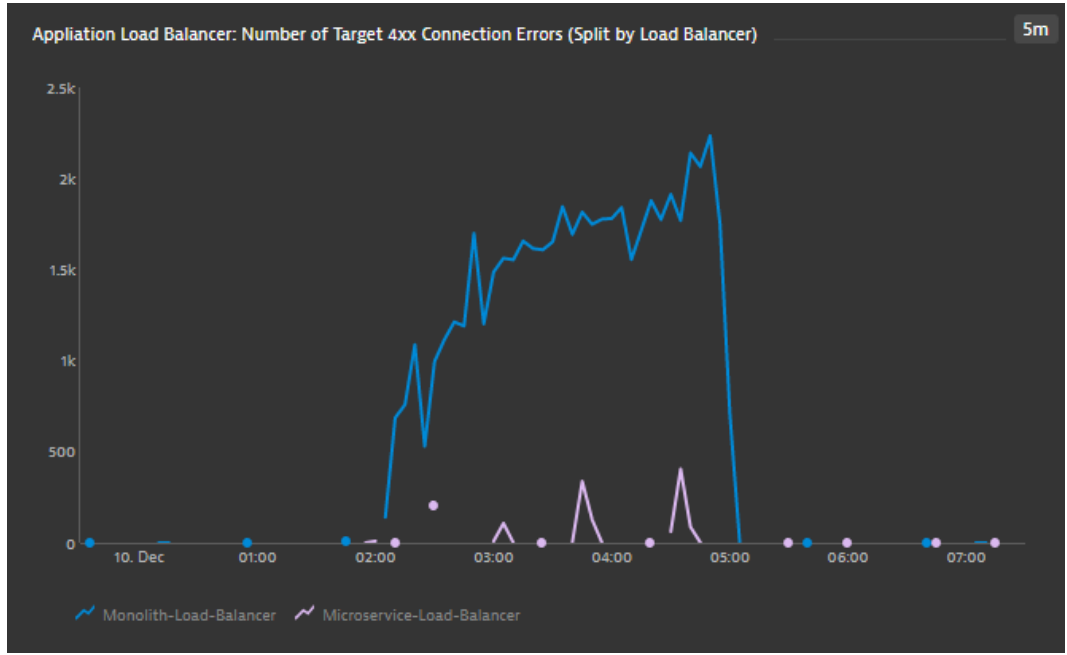

Figure 15: Total number of 4xx errors, split by load balancer

In Figure 15, we can see the number of 4xx errors split by load balancer. While there are periods where some 4xx errors occur for the microservice architecture, these are short, isolated periods. The monolithic architecture sees a steady climb in 4xx errors after 2 am until 5 am, when demand on the application was the highest. While errors are not a good thing, the server is still sending a message back saying the requests were received means that the application was not down, which is consistent with other perspectives that have yet to be analyzed.

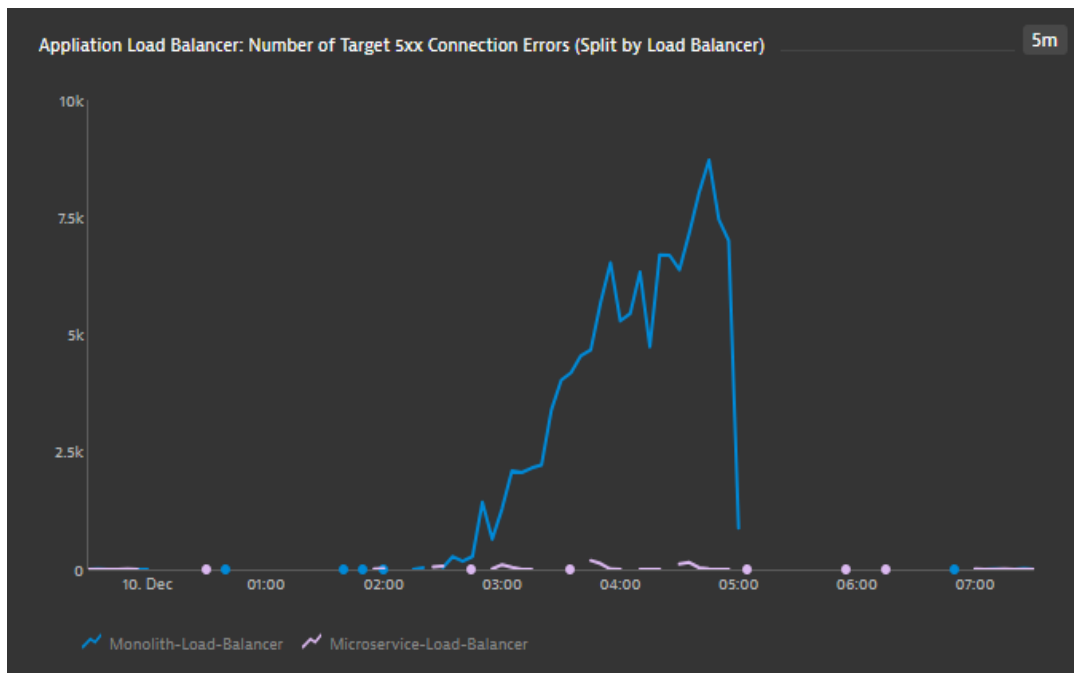

Figure 16: Total number of 5xx errors, split by load balancer

Unlike 4xx errors, which can occur for organic reasons and in low quantities, they may not need to be investigated. Any amount of 5xx errors is considered something not to be ignored. Small quantities of these errors are shown in Figure 16 for short period microservice architecture. In contrast, the monolithic architecture sees these spikes as higher and higher as the load on the application increases. The server cannot send any response back to the simulated users coming from the load generators.

D. CPU Utilization

CPU is an important dimension when trying to factor in the allocation of resources. It becomes a complex resource that can be allocated and used in several different ways within a datacenter. For instance, it is critical to know the software requirements that are to be run for clock speed, number of cores, memory cache values, and even the register size of the processor. Other important factors include whether an application will be running on a dedicated machine versus a virtual machine, where virtual resources can be changed or modified later as required.

Per the proposed methodology, CPU utilization percentage was the determining factor when either application cluster was determined to scale up or down in the total number of cluster nodes.

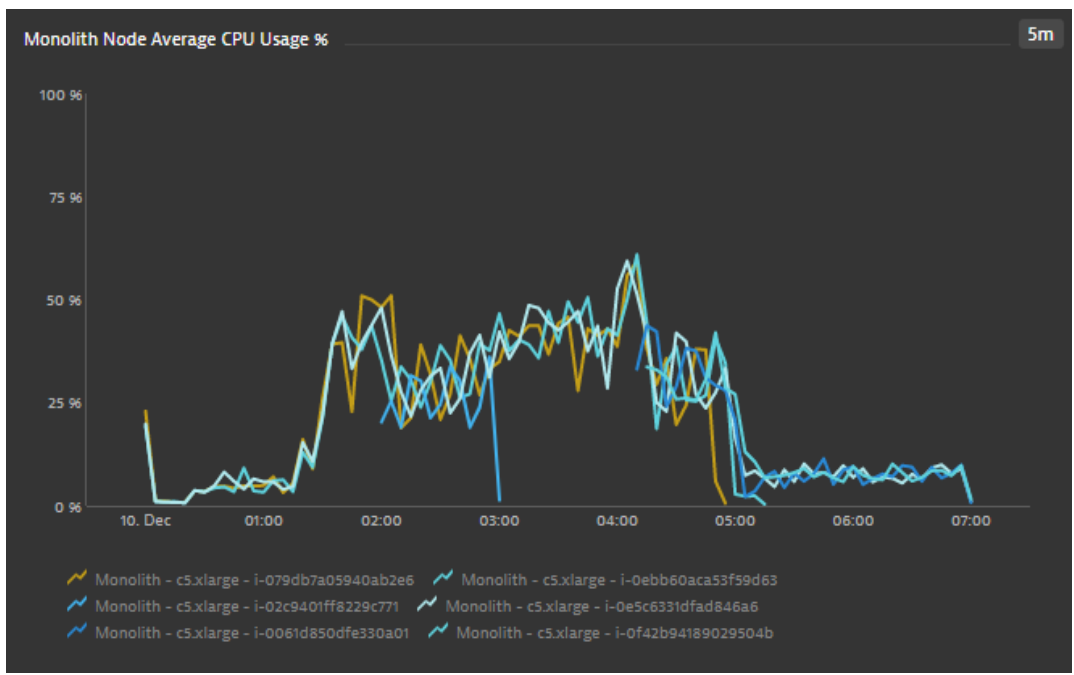

Figure 17: Average CPU utilization per node, monolithic architecture

Figure 17 shows that the average CPU utilization is shown broken apart per node instance in the monolithic cluster.

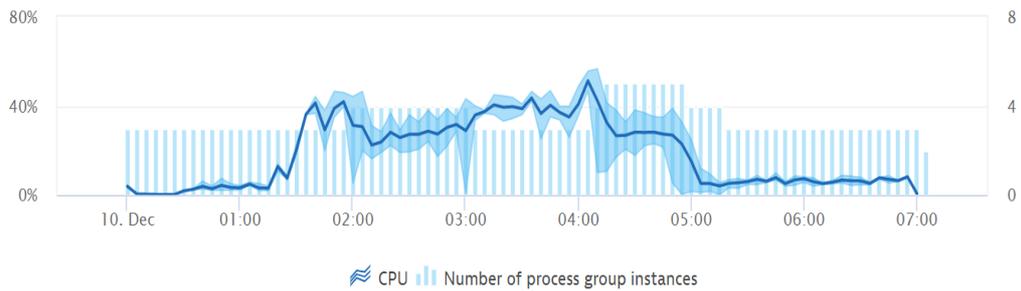

Figure 18: Average CPU utilization of EasyTravel Frontend of all instances (Monolith)

In Figure 18, the bars represent the average of all running monolithic cluster frontend instances at a given time. In contrast, the line represents the average CPU utilization of the monolithic version of the EasyTravel application Frontend responsible for most application-related CPU utilization on each cluster node. One thing that becomes immediately obvious in either in (Figure 17 or 18) is that the average utilization does not cross our threshold to scale the cluster (50%). The application's throughput suffered even though it was shown in the previous sections after a certain amount of load was generated. In Section 3.1, it was expected that the monolithic cluster would not scale as efficiently as the microservice-based cluster because each monolithic node would only have one running instance of all the application's different components. This bottleneck was expected to stretch the node instance(s) resources and trigger the node management technique set to keep the cluster utilization of either architecture to fifty percent overall utilization and then add or remove cluster nodes to achieve that goal. As the monolithic cluster nodes, average utilization approached fifty percent, the application's overall throughput dropped, and the cluster could no longer handle the number of requests it received. There are two instances around 2 am, and 4:15 am where the CPU utilization criteria did add in an additional cluster node at 2 am, and at 4:15 am, there were two additional cluster nodes added. We can see slight increases in the application's throughput at both points, but the new nodes immediately become just as overwhelmed as the already running nodes. While explored in greater detail in a subsequent Section (4.2.6), the nodes did not spike higher in CPU utilization and triggered additional cluster nodes to be added as a result of the garbage collection of objects.

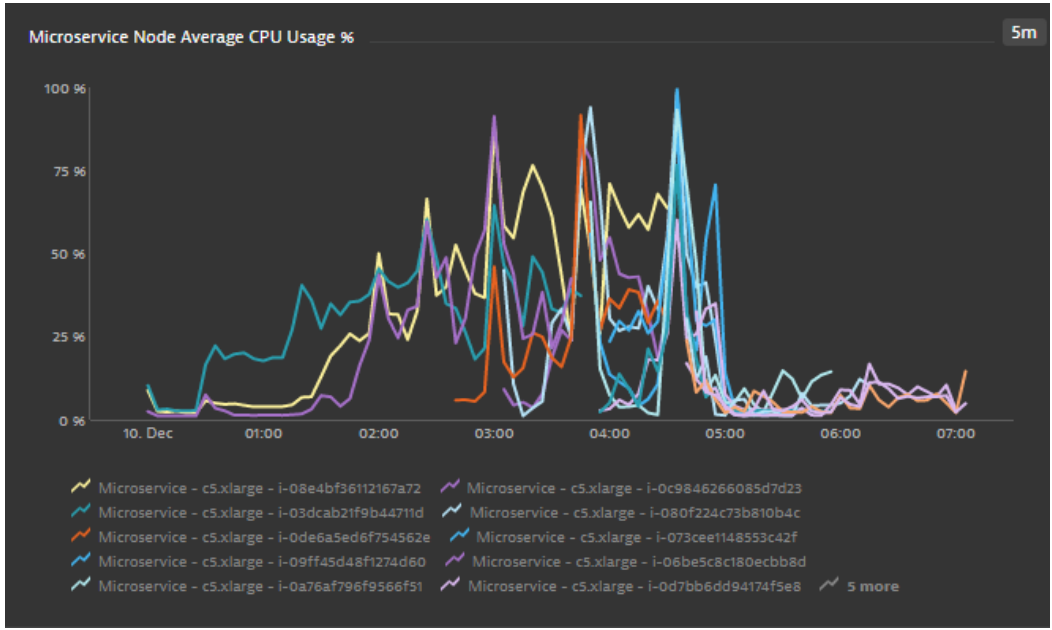

Figure 19: Average CPU utilization per node, microservice architecture

Figure 19 shows that the average CPU utilization is broken apart per node instance in the microservice cluster.

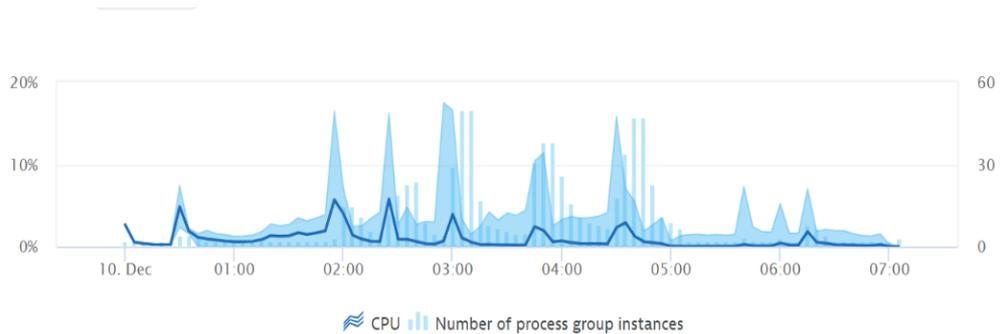

Figure 20: Average CPU utilization of EasyTravel Frontend all instances (Microservices)

In Figure 20, the bars represent the number of running microservice cluster frontend instances during the timeframe. In contrast, the line represents the average CPU utilization of the microservice version of the EasyTravel application. A few things immediately stand out when compared to the monolithic architecture representations from Section 3.2 Design. First, the average CPU utilization per frontend instance is significantly lower than its monolithic counterpart. Second, it can be seen that the total number of running frontend instances is far greater than the monolithic counterpart. This change in the number of instances is quite intuitive and was to be expected. The microservice architecture splits the application into three separate containers: the frontend, backend, and database. These components run their own containerized services. Each can be scaled to have a more significant number of instances running on each microservice cluster node per the values set up in the methodology to produce additional supply when demand increases. This is in stark contrast to the monolithic architecture, which only can scale by adding in additional cluster nodes. The overall CPU of the frontend instances for the microservice architecture remains relatively low overall. In contrast, the number of instances approaches close to 60 at times. In contrast, the maximum number of frontend instances for the monolithic architecture is six since it is bound to the number of instances running is constrained by the total number of running monolithic cluster nodes.

Another stark contrast seen in the microservice-based architecture versus the monolithic architecture is that the overall cluster node CPU utilization keeps rising with increased load and demand put on the application. As hypothesized, this results in the scaling of overall cluster nodes' number to increase as expected in the methodology. The issues and limitations of the monolithic version of the architecture are overcome with the microservice architecture by adding additional front-end capacity on a cluster node before the running containers cannot handle the requests.

### E. Availability/Operation

Perhaps one of the most important metrics to consider when testing or analyzing applications is availability. Application availability is the measure in which an application is determined to be available and operational/functional and used to fulfill or complete its purpose. It is one thing for an application to load, but if a user can browse a site like Amazon.com yet cannot checkout, that is a significant issue. In Section 2.2, the analysis which was performed done looked at some internal server metrics such as the throughput of the application and the architectures' overall performance response time. Here, the analysis will look at some of the synthetic testing set outside of the AWS availability zone where all the testing occurred. Since the load balancer for each architecture was publicly available over the internet, a set of synthetic monitors was set up to monitor each architecture homepage's availability. If it could be loaded, the load time lengths and some other load times were, and some other metrics about its overall performance from six different datacenter locations in the United States, as shown in Figure 21. These tests were set up to run from Oregon, Chicago, Cheyenne, Los Angeles, South Carolina, and Texas.

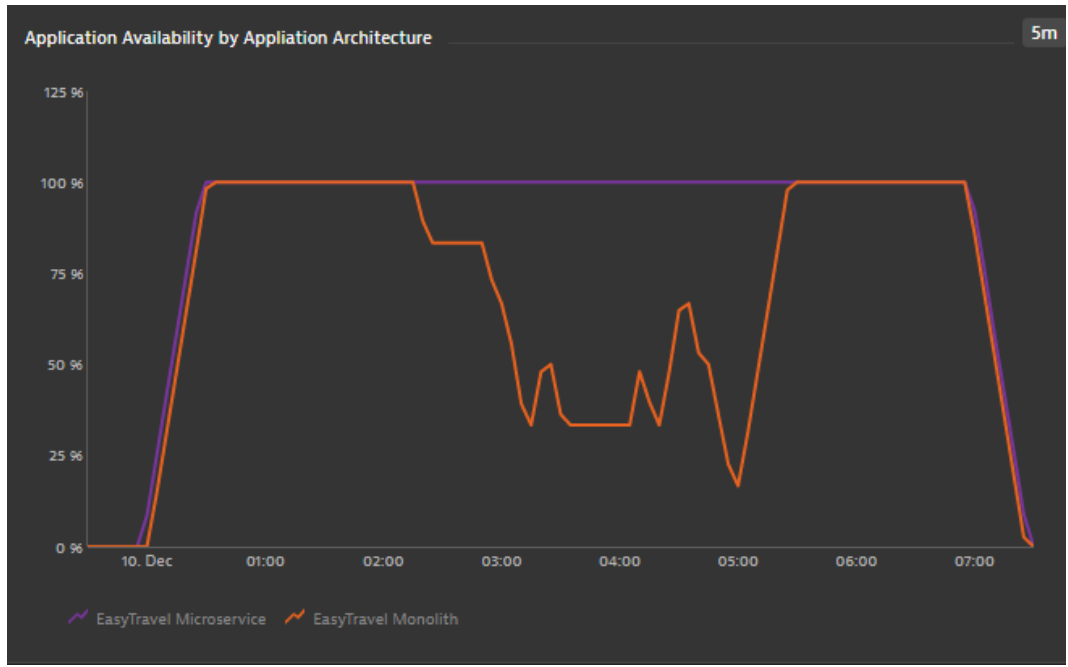

Figure 21: Application availability (split by application architecture)

Here in Figure 21, the availability is shown as a percentage for both application architectures. As with the previous sections, once the monolithic architecture starts to have problems, it is again seen to reduce the site's overall availability. The microservice instance remains fully available throughout the duration of the test.

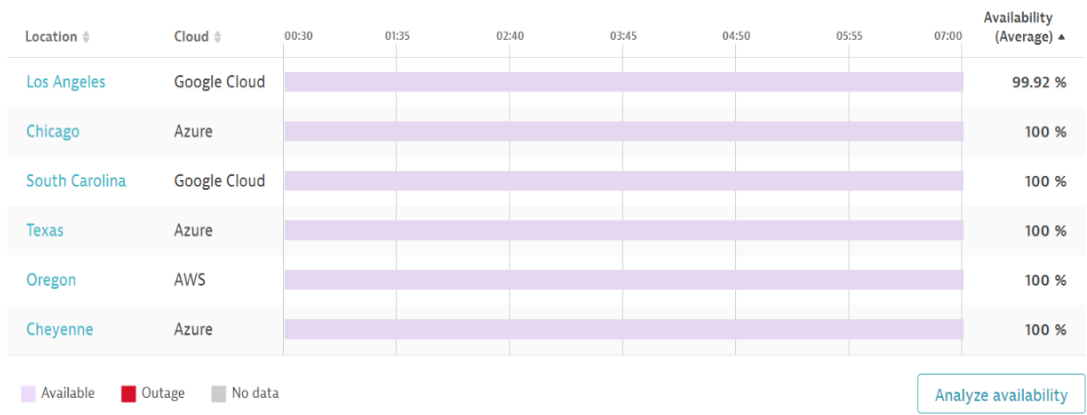

Figure 22: Application availability split by geography of synthetic tests (Microservice)

Figure 22 shows the availability has been broken apart by location/datacenter. Also visualized are generalized statistics around downtime, visually complete (a point-in-time metric that measures when the visual area of a page has finished loading), and total load duration, to name a few, for the microservice cluster.

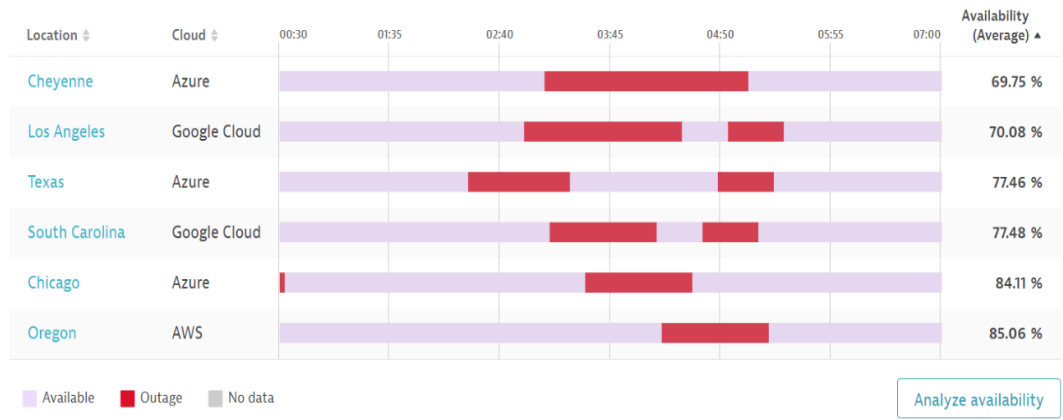

Figure 23: Application availability split by geography of synthetic tests (Monolith)

Looking at the availability graphs broken apart by location/datacenter for the monolithic architecture in Figure 23, it is much more interesting to dissect and interpret. It is seen as different overall with stats such as downtime, visually complete, and total load duration. However, when looking at the downtime for the individual locations represented as the bar graphs' red sections per location, the downtime is not concentrated at certain times; otherwise, all locations would have the outage graphs lined up with one another. These non-overlapping sections show it can be reached from one of many locations meaning that despite the monolithic architecture being less responsive and available overall, it was still processing transactions the entire duration of the test. However, if a given transaction was successful or failed, it was a function of the system's resources.

F. Garbage Collection/Suspension Time

In the interest of brevity for this section, only some very cursory and brief overviews of garbage collection will be explained. This section will present how it affected the application testing performed and its negative impact on the application and virtual machine performance. To fully understand topics covered in this section more thoroughly and their importance to the results of the testing done for this work, it is recommended to the reader to read into other papers and articles exclusively about garbage collection in detail. It is also expected that the reader has fundamental knowledge of modern computer architecture, including CPU, memory, and concepts of how objects are created, modified, and destroyed by any of the following: the host OS, guest OS, containers, and applications running on any prior mentioned systems. In essence, garbage collection in the context of computer architecture refers to how objects in memory are managed once they are no longer in use by any program. This essentially means how to destroy these objects to free up computer memory for re-use. There are many ways garbage collection can be handled, and there is no one size fits all solution for all applications. What may be optimal for one application could very well be inefficient for another. When a poor garbage collection strategy is used, it can cause considerable performance problems for an application. Regardless of the garbage collection strategy used, there will be a period on the CPU where the application processing must be suspended, and garbage collection occurs. This stop in CPU is referred to simply as suspension. How long garbage collection occurs is referred to as garbage collection time and is expressed as a value of time, whereas suspension is expressed as a percentage. As an analogy, consider if you and other drivers drive down a straight multilane road with traffic lights along with it. You can think of the number of times traffic had to stop at a light rather than being able to go straight through, as your suspension percentage, and the amount of time you were stopped at the light as your garbage collection time. All traffic, regardless of the number of lanes, must follow these rules, which would be cores or threads of a CPU for this analogy. The more times you stop or, the longer you are stopped, the more time it takes to get where you are going, or in a computer's case, the worse the application's performance.

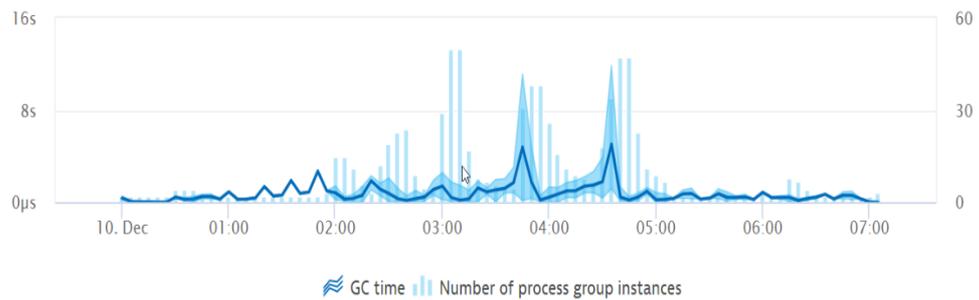

Figure 24: Garbage collection time vs. number of microservice frontend instances

In Figure 24, the garbage collection time for the microservice-based application, the garbage collection time per interval is relatively low due to having a much higher number of instances (177 with microservices architecture versus 6 with monolithic architecture) to spread the workaround. At the peak times during the highest loads, it was only a few seconds.

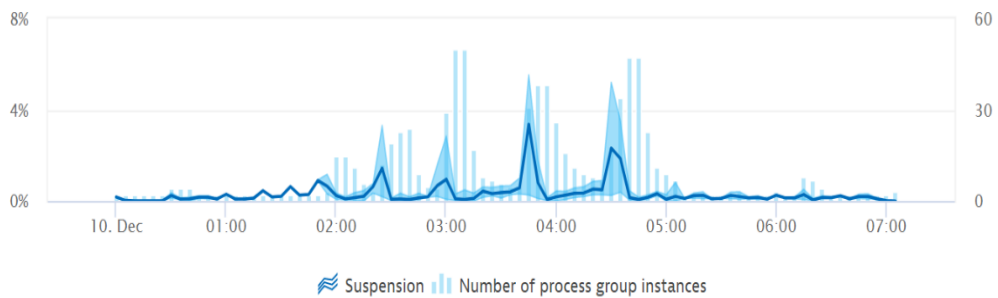

Figure 25: Suspension % vs. the number of microservice frontend instances

Similarly, in Figure 25, the percentage of time spent in garbage collection (suspension) is less than one percent, and at the peak, loads never went higher than four percent since the microservice architecture had many more instances to spread the workaround than the monolithic architecture: 177 versus 6 respectively.

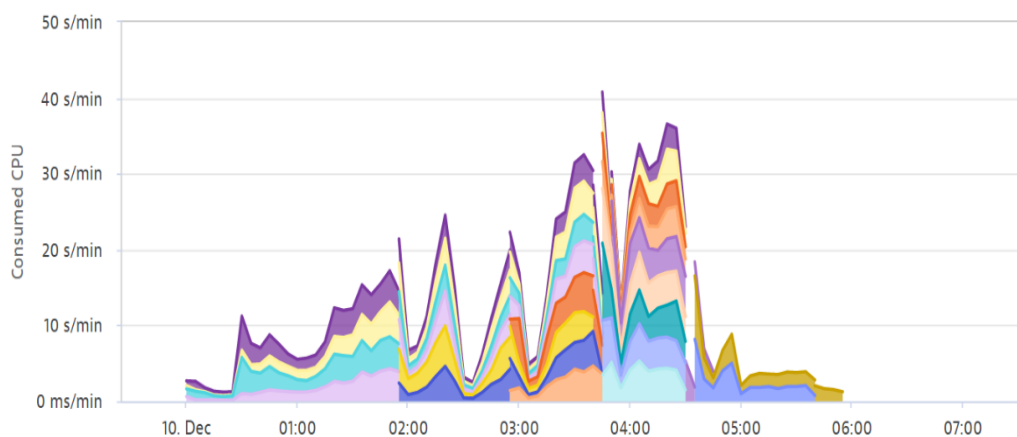

Figure 26: CPU consumption per service (Microservices)

Figure 26 shows when the service is spent executing on a CPU over time. Due to limitations in how the data can be exported, there is little context given. The critical thing in this graph is that the areas that took up the most time on the CPU in the microservices architecture were background services and the services required to run the application, such as the frontend component. The garbage collection time was very little overall time compared to everything else.

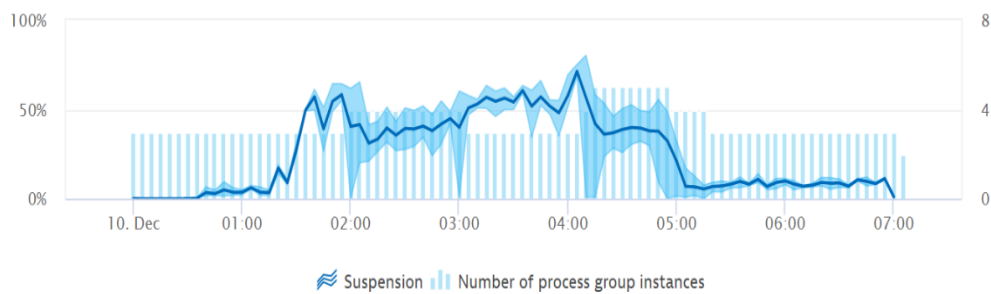

Figure 27: Suspension % vs. the number of monolith frontend instances

Figure 27, a stark contrast to the microservice suspension time, shows that the application is being stopped half of the time for garbage collection to take place. Remember, all cores are suspended when garbage collection is taking place. Since the cores are being suspended, this does not directly affect our overall CPU utilization numbers. Our test methodology uses CPU utilization to increase the number of monolithic cluster nodes based on CPU utilization. This suspension is the primary reason why the monolithic cluster did not scale the number of nodes up as the load increased in the predicted way.

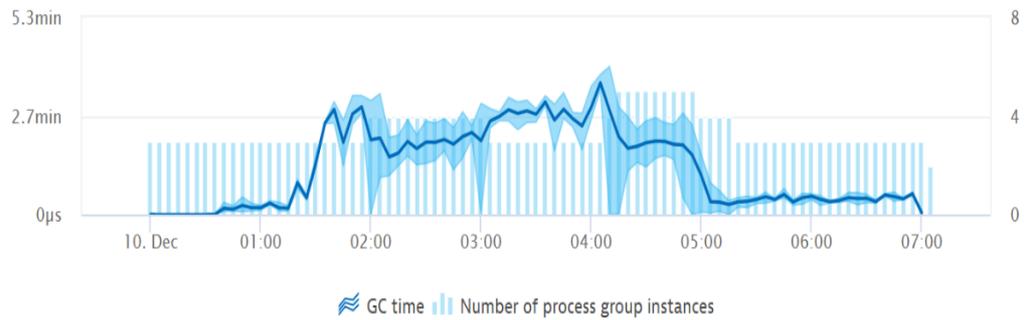

Figure 28: Garbage collection time vs. number of monolith frontend instances

In Figure 28, garbage collection time in the monolithic instance is predictably much higher than what was seen in the microservice architecture, measured in minutes rather than seconds. This can be attributed to having many more instances to spread the workaround.

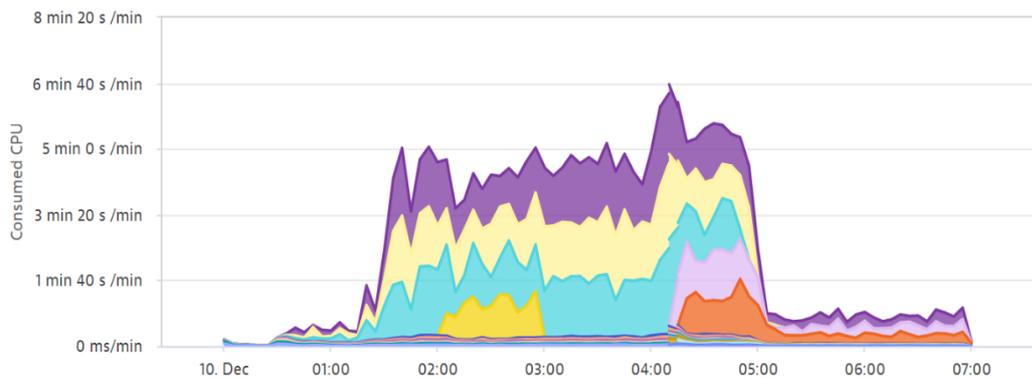

Figure 29: CPU consumption per service (Monolith)

Coming now to Figure 29 for the monolith architecture, it can be seen in the graph the amount of time a service is spent executing on a CPU over time is expressed in minutes rather than seconds as in the microservice architecture. The reason for including this visualization is that the tool used to capture this data can hide time spent in background tasks and garbage collection time. While this makes no significant difference in the microservice architecture, there is a significant change when the same CPU utilization graph is displayed with the garbage collection time removed.

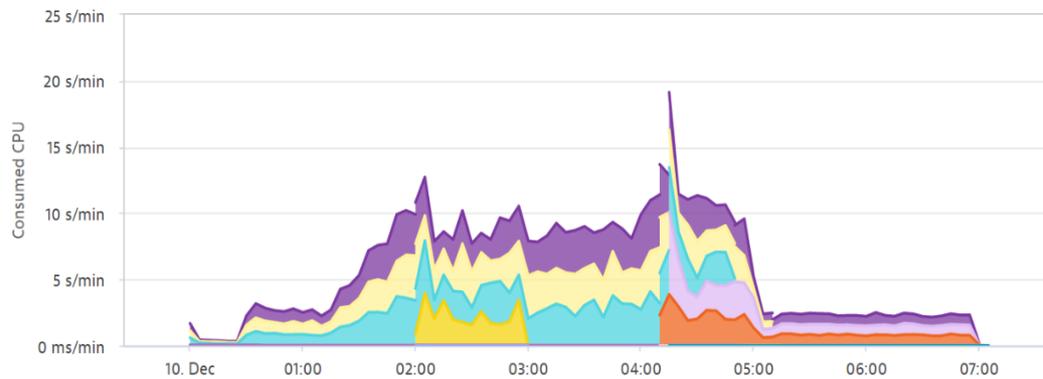

Figure 30: CPU consumption per service, background/garbage collection usage removed (Monolith)

In Figure 30, the monolithic architecture with the background/garbage collection CPU usage hidden from the application components is now measured in seconds. It is similar to what is shown in the same visualization of the microservice architecture (Figure 26). This difference is likely due to how the application must create, then destroy the object in memory before handling more transactions. The microservice architecture most likely does not have this same bottleneck due to its ability to create more instances to handle more transactions. Even if the system uses the same garbage collection strategy, the container operating the single service is suspended during garbage collection rather than the entire application.

G. Application Scaling

As mentioned in this section, the monolithic scaling did not operate as intended per the methodology. It was theorized that the monolithic architecture would not handle as many transactions per cluster node due to its inability to increase the capacity of specific services needed on the same host or hosts within the same cluster as the microservice-based architecture. As a result of this, the monolithic cluster's expected outcome would have a higher node count for the same or a smaller number of transactions than its microservice counterpart. With the garbage collection issues, the monolithic architecture experienced keeping the CPU threshold just below the level of autoscaling. The implementation ended up showing something different than was predicted.

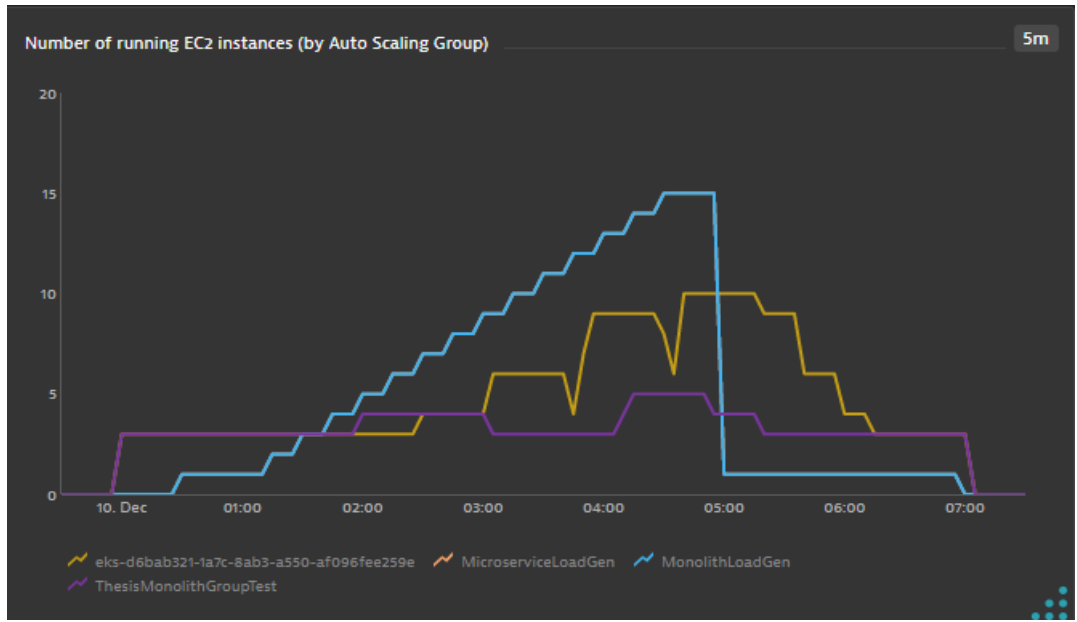

Figure 31: Application scaling split by autoscaling group

In Figure 31, one can observe the cluster node count of all the running nodes for each cluster. The microservice and monolithic load generation cluster scale identically, so the light blue line paints overtop the orange node count for the microservice load generators. Until around 2 am the lines for the microservice application cluster (yellow) and the monolithic application cluster (purple) also paint overtop one another as they are the same. Just after 2 am, the monolithic cluster increases its node count from three to four, yet about an hour later, the average CPU of the cluster falls below the threshold. It drops back down to three for a while before adding and removing a couple of nodes around the test's peak load times. The microservice application cluster behaves precisely as predicted, with both the number of cluster nodes and a number of microservice instances rising as needed when the load generation increases over time.

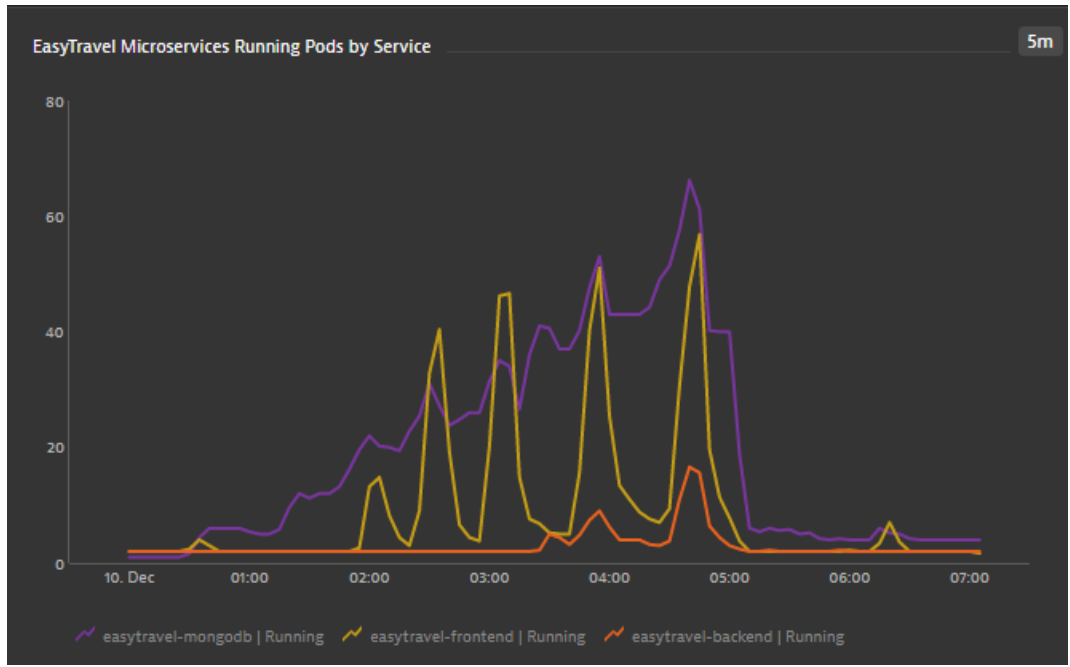

Figure 32: EasyTravel microservice pod scaling

In Figure 32, the Pod scaling within the microservice application cluster reveals some fascinating things about the scaling behavior and perhaps its application. Initially, the Kubernetes horizontal Pod autoscaler in EKS was set to have one instance of the database running and two instances of each of the frontend and backend components EasyTravel. Like the cluster itself, criteria around CPU utilization of the Pods were put in place to scale the number of Pods to keep the average CPU target of all the Pods of any given type at 80%. The CPU value was set much higher here than the cluster nodes because of the speed at which more Pods can be created. Since there is no virtual or physical machine to turn on, load dependencies give a ramp-up/warm-up time for then as long as there is capacity within the cluster for a Pod to be created; it is nearly instantaneous, and works can begin immediately. The least noteworthy thing to be seen here is the backend service, which, until the loads started to reach the peak, never increased, and even when they did increase, it was not as drastic as the other two services. The database service seemed to scale very linearly as the load increased, so did the database Pod instance. Finally, and by far, the most interesting, is the frontend service scaling. The frontend Pods, for the most part, stay about the same. They spike up very high on several occasions and stay that way for around twenty minutes before scaling back down. This data could not be correlated where this spike in frontend service Pod count occurs to any other metric. It does not follow load generation, which is much more linear in which the EasyTravel database Pod instance count seems to have a much stronger correlation. There are also no spikes in response time, requests, errors, or other metrics that suggest a need for more running instances of that type of Pod. Out of all the data gathered, the biggest question left unanswered is what produced that behavior.

# 5. Conclusions and Future Work

This paper examined how a single application architecture converted from a monolithic codebase architecture to a containerized service-based architecture allows for better resource utilization and scalability. The scale, scope, and dimensions examined in this work differ from any other work or approach reviewed. Undoubtedly, there are challenges and costs in refactoring an application to be used in a modern containerized orchestration environment like Kubernetes. These costs and challenges are mostly upfront and do not outweigh the containerized service-based architecture's benefits.

The ability for parts of an application to be split up from one to many and then scale out additional capacity on the same or another host within a clustered environment as demand shifts and changes without manual intervention is much better suited for the ever-changing digital landscapes of today. The benefits do not start once the application has reached run time but through the entire application life cycle. An application made of Application Programming Interfaces (APIs) can be broken down into tightly coupled code with well-defined interfaces interacting with all services. This decoupling and reusability allow services to be reused among multiple applications and faster code changes and deployments since only individual services are being modified, not the entire application.

The data presented showed how different services, when using a containerized service-based architecture, will dynamically scale and how different scaling can look based on the type of service. Just because the load was added linearly did not mean that each service scaled the number of containers for all services linearly. What is shown in this paper shows only the start of what is possible. Managing many applications using the approach and technique outlined in this paper could increase performance, and productivity, speed up innovation and reduce the total infrastructure cost of virtually any organization with all but the smallest IT footprints.

For future work, increasing the scope from just a single application to multiple interconnected applications that interact with one another would greatly expand the ability to show how datacenter resources are better utilized and scaled in a containerized service-based architecture. The number of overall nodes in each cluster for a monolithic application and one that comprises microservices may not differ significantly for a single application. Due to industry shifts towards containerization as the preferred architecture, it would be hypothesized that the overall number of cluster nodes needed to support a complete implementation of multiple applications would be less than that of its monolithic counterpart deployed across multiple virtual machines. This type of work would produce valuable results if the scope were broad enough to say what average cost savings would be for an organization to refactor/redevelop the organization's applications.

**Acknowledgments**

The authors would like to thank anonymous referees for their valuable suggestions and comments.

**Abbreviations**

OS: Operating System; SOA: Service-Oriented Architecture; B2B Application Programming Interface: Business Application Programming Interface; I/O: Input/Output; CPUs: Central Processing Units; RAM: Random Access Memory; GPUs: Graphics Processing Units; SANs: Storage Area Networks; APIs: Application Programming Interfaces; NAS: Network Attached Storage; AWS: Amazon's Web Services; AWS EC2: Amazon Web Services Elastic Compute Cloud; B2B: Business-to-Business; UI: User Interface; AMI: Amazon Machine Image; EKS: Elastic Kubernetes Service; URL: Uniform Resource Locator; XL: Extra Large.